\documentclass[11pt]{article} 

\usepackage{amsmath,amsthm,latexsym,amssymb,amsfonts,epsfig}

\newtheorem{Theorem}{Theorem}[section]
\newtheorem{Definition}{Definition}[section]
\newtheorem{Lemma}{Lemma}[section]
\newtheorem{Corollary}{Corollary}[section]

\newtheorem{Example}{Example}[section]

\newcommand{\be}{\begin{equation}}
\newcommand{\ee}{\end{equation}}
\newcommand{\ba}{\begin{eqnarray}}
\newcommand{\ea}{\end{eqnarray}}

\title{{\sf Quantum gravity in the triangular gauge}} 
\author{
{\sf T. Thiemann}$^1$\thanks{{\sf 
thomas.thiemann@gravity.fau.de}}\\
\\
{\sf $^1$ Inst. for Quantum Gravity, FAU Erlangen -- N\"urnberg,}\\
{\sf Staudtstr. 7, 91058 Erlangen, Germany}\\
}
\date{{\small\sf \today}}

\makeatletter
\@addtoreset{equation}{section}
\makeatother

\begin{document} 

\maketitle

{\sf

\begin{abstract}
Vielbeins are necessary when coupling General Relativity (GR) to fermionic 
matter. This enhances the gauge group of GR to include local Lorentz
transformations. 

In view of a reduced phase space formulation of quantum gravity, in
this work we completely gauge fix that Lorentz gauge symmetry by 
using a so-called triangular gauge. 
Having solved the Gauss constraints already classically opens access 
to new Hilbert space representations which are free of the complications 
that otherwise arise due to a non Abelian gauge symmetry. 

In that sense,
a connection formulation as being pursued in Loop Quantum Gravity is 
no longer the only practicable option and other less dimension 
dependent representations 
e.g. based on triads and even metrics suggest themselves. 
These formulations make it 
easier to identify states representing non-degenerate quantum geometries 
and thus to investigate the hypersurface deformation algebra which 
implicitly assumes non-degeneracy.      
\end{abstract}

\section{Introduction}
\label{s1}

As Cartan showed \cite{1}, there are no multivalued (i.e. spinor) 
representations of 
the diffeomorphism group of a manifold on $M$. While in (asymptotically)
flat spacetime 
Lorentz transformations are possible diffeomorphisms which do 
carry multi-valued represntations, these cannot 
extended to general diffeomorphisms. Accordingly 
one has to separate between diffeomorphisms and (local) Lorentz transformations
in order to accomodate spinors (possible 
on manifolds that carry a spin structure). 
This can be accomplished by objects that
carry both representations of the Lorentz group  
and representations of the diffeomorphism 
group in order to couple fermionic matter to geometry in a diffeomorphism 
covariant fashion. These are the Vielbeins which are at the same 
vector fields on $M$, transform in the defining representation of the
Lorentz group and locally diagonalise the spacetime metric 
to the Minkwoski metric. Combining them with Dirac matrices which intertwine 
tensor 
products of two 
double valued spinorial representations and the single valued 
defining representation one obtains a covariant description of spinors
which reduces to usual one in the the flat spacetime limit.

As fermions are the fundamental building blocks of matter, general 
relativity (GR) must be described in terms of Vielbeins rather 
than metrics. This leads to the Palatini formulation  of the 
Einstein-Hilbert action and one may adopt a first or second order 
approach which are equivalent in the vacuum case. In the first 
order approach, one introduces in addition to the Vielbein a Lorentz 
group connection as an independent field while in the second order 
approach that connection is defined as the torsion free 
spin connection of the Vielbein from the outset. In the vacuum case the 
Euler Lagrange equations of the first order formulation 
fix the connection to be that spin connection, in the presence of fermionic 
matter an additional torsion term bilinear in the fermion field arises.
As the action is quadratic in the connection, eliminating the connection 
by its equation of motion from the action yields a 4-fermion interaction 
term which is absent in the second order formulation.
 
Whether one uses the first or second order formulation, one is lead 
to a formulation of GR coupled to (fermionic) matter in terms Vielbeins
and an enhanced gauge symmetry which includes diffeomorphisms and 
local Lorentz transformations. In $D+1$ spacetime dimensions, the Vielbein 
carries 
$(D+1)^2$ degrees of freedom and 
and there are $\frac{1}{2}\;D(D+1)$ local Lorentz constraints which means 
that there are $(D+1)(D+2)/2$ Lorentz invariant degrees of freedom 
which are precisely the number of independent componnents of the spacetime
metric.
  
In the canonical approach these symmetries are formulated in terms 
of constraints, i.e. functions on the canonical phase space whose 
Hamiltonian flow generates spacetime diffeomorphisms and Lorentz 
transformations. Observables of the theory are phase space functions 
that are invariant under those local gauge symmetries. The fundamental 
building blocks of geometry are thus e.g. the co-Vielbein $e$ 
(a one form with values in the representation space of the
defining representation of the Lorentz group) and its conjugate 
momentum $P$ (a vector field density of weight one with values 
in the representation space of the contragredient representation).
Here $e$ captures information about the intrinsic D metric $q$ of the 
Cauchy surface underlying the canonical framework while $P$ encodes 
information about its extrinsic curvature $k$.
When the metric defined by the Vielbein is non-degenerate
an equivalent, dual description is in terms of variables $E$ and 
$K$ with opposite assignments of tensorial and Lorentz representation 
character.

In the canonical approach it is customary to gauge fix the boost part of 
the Lorentz group using the so called time gauge \cite{3}. This leaves 
the rotations unfixed and leaves the corresponding Gauss constraints 
of the rotation subgroup intact. One motivation to get rid of the boosts is 
that the Lorentz group is not compact while the rotation group is 
compact. In quantisations of the theory, most prominently 
in Loop Quantum Gravity (LQG) \cite{4}, one so far has not been able to 
cope with the non-compact gauge group case while for the compact case 
proper Hilbert space representations of the canonical commutation 
relations and the local rotations have been found \cite{5}. 

As one solves the corresponding Gauss constraints at the quantum level
it is natural to formulate the theory in terms of connections 
\cite{6} $A$ rather than $P,e$ above because connections transform 
covariantly under rotations when integrated along path which is why they 
are so popular in lattice gauge theory \cite{7} for compact gauge groups 
where one employs gauge invariant 
Wilson loop functionals, a fact from which LQG derives its name.

While the formulation in terms of connections opens access to the 
arsenal of techniques of lattice gauge theory and an entirely new 
formulation of its dynamics, the fact that one still is dealing with 
a non-Abelian, non observable gauge symmetry is mathematically 
challenging. In particular, while it is possible to actually 
solve the Gauss constraints in closed form in terms of spin network 
functions \cite{8}, the operators that encode the quantum geometry 
are very complicated to deal with. Even though in the classical theory 
the Poisson brackets say between the $E$ variables vanish, in 
the quantum theory the cooresponding operators 
do not \cite{9}. This is no contradiction 
because what one quantises is not functions on the phase space but 
their Hamiltonian vector fields on a space of functions of the connection
and these are not commuting even in the classical theory. Yet this 
fact implies that many operators in the quantum theory have rather 
complicated spectra. For instance, the spectrum of 
volume operator \cite{10} 
which is pivotal for the quantum dynamics \cite{11} is not accessible 
in closed form and this fact, among others, has decelerated progress 
in the field.

In this paper we ask the natural question whether it is not conceivable 
and more convenient to also gauge fix the rotation part of the 
local Lorentz group, especially in view of recent approaches to gauge 
fix the spatial diffeomorphism and Hamiltonian constraints of 
the theory using material reference systems \cite{12} which has the 
advantage of sidestepping a huge number of complicated tasks 
(densely defining and solving 
the quantum constraints, construct a physical inner product, construct 
the quantum observables). We show that there exists indeed a 
family of gauges, called {\it triangular gauges} which have the 
following convenient properties:\\
1. They can always be installed, no matter where on the constraint 
surface one starts form.\\
2. There are no Gribov copies, i.e. the gauge condition intersects 
each gauge orbit precisely once.\\
3. At least when the classical geometry is non-degenerate, one can solve 
uniquely 
for a set of momenta which are canonically conjugate to the configuration
variables that one fixed.\\
4. The remaining spatial diffeomorphism and Hamiltonian constraints can 
be explicitly and locally expressed in terms of the unfixed configuration and 
unsolved momentum variables, no non-local integrations are necessary.\\
5. The gauge fixed phase space can alternatively 
be formulated in terms of 
1-forms and (D-1) forms or D-metrics and dual tensors 
which is of advantage in the quantum theory.\\
6. In particular, the non-commutativity 
issues of \cite{9} disappear and the mathematics of the theory
simplifies, in fact one has access to a whole new class of Hilbert space 
representations which ease the investigation of the issue of quantum metric 
degeneracy \cite{13}. In that sense, the triangular gauge reduction 
is similar in spirit with Quantum Reduced LQG (QRLQG) \cite{13a} in 3d where 
one imposes a diagonal gauge condition on the triad and in that sense also 
reduces the spatial diffeomorphism constraint. The advantage of the 
triangular gauge is that one can show that it is a perfect gauge in the 
the above sense and 
in particular reduces the theory to the ADM phase space in any dimension
as far as the geometrical degrees of freedom are concerned.
It can therefore either be reduced further in a complete reduced phase space 
quantisation approach or be used in operator constraint quantisation of 
full GR with spatial diffeomorphism and Hamiltonian constraint intact.\\  
\\
\\
The architecture of this paper is as follows:\\
\\

In section \ref{s2} we briefly recall the canonical formulation of GR in 
terms of Vielbein variables and conjugate momenta in $D+1$ spacetime 
dimensions with the time gauge to fix boosts installed but with local 
rotation gauge symmetry still intact.

In section \ref{s3} we introduce the triangular gauge and explore its 
properties.

In section \ref{s4} we carry out the symplectic reduction with respect  
to the upper triangular gauge, i.e. we solve the Gauss constraint classically
and write the constraints in the reduced variables. 

In section \ref{s5} we construct new Hilbert space representations of 
canonical GR in the time gauge and point out the advantages and 
disadvantes as compared to the non gauge fixed formulation. For instance,
spectra of geometrical operators can be computed in closed form.
In particular we discuss
convenient properties of the upper triangular polarisation in the 
canononical quantisation of the Hamiltonian constraint both in the full
theory and in its cosmological truncation.

In section \ref{s6} we summarise and conclude. 

In the appendix we include some background material on non-degenerate 
states in particular but not exclusively within the current LQG representation.

\section{Canonical GR with Vielbeins in the time gauge}
\label{s2}

To arrive at the Vielbein formulation one may start from the Palatini 
action based on the Lagrangian \\
$\det(\{e_\rho^K\})\;F_{\mu\nu}\;^{IJ}\;e^\mu_I\;e^\nu_J$ 
with spacetime Vielbein $e^\mu_I,\;I,\mu=0,..,D$ and curvature 
$F$ of the connection $A$ which is kept independent in the first order 
formulation or is chosen as the spin connection of the spacetime 
Vielbein in the second order formulation. The detailed Legendre
transformation of this action including (fermionic) matter terms can be done
\cite{14} and leads to primary and secondary, first class and second class 
constraints. One gets rid of the boost part of the Lorentz constraints
by using the time gauge $e^0_j=0$ with $j=1,..,D$ and keeps the rotation 
constraints. Here we will not go through that lengthy derivation but rather
proceed via extending the phase space of the ADM formulation 
\cite{15} and introducing the Gauss constraints which leads to the same 
final result. This will also introduce our notation:\\
\\
We work in $D$ spatial dimensions with a manifold $\sigma$ which 
is diffeomorphic to the leaves of the foliation of the spacetime manifold
by Cauchy surfaces of a globally hyperbolic spacetime. Tensorial indices
will be denoted $a,b,c,..=1,..,D$. The intrinsic metric of $\sigma$ 
is denoted by $q_{ab}$. We introduce a spatial co-D-bein $e_a^j$ where  
$j,k,l,..=1,..,D$ denote the components of $e_a$ with respect to some 
basis of the representation space of the the definig reprsentation of 
SO(D). Then 
$q_{ab}=\delta_{jk}\; e^j_a\; e^k_b$ is invariant under local 
SO(D) rotations of the Vielbein. 
Tensorial indices $a,b,c,..$ will be
moved in what follows by $q_{ab},q^{ab}$ with $q_{ac}\; q^{cb}=\delta_a^b$,
representation indices $j,k,l,..$ will
moved in what follows by $\delta_{jk},\delta^{jk}$ 
with $\delta_{jl}\; \delta^{lk}=\delta_j^k$.

Let $\Gamma_{ab}^c=\Gamma_{ba}^c$ be the torsion free 
Levi-Civita connection compatible with $q$ and 
$\omega_{ajk}=-\omega_{akj}$ the so(D) valued 
spin connection of $e$. Then the covariant 
differential $D$ is defined on arbitrary mixed tensor fields by 
\be \label{2.1}
D_a\; T_{b..j..}^{c..k..}=
\partial_a T_{b..j..}^{c..k..}
-\Gamma_{ab}^d\;T_{d..j..}^{c..k..}
-..
+\omega_{aj}\;^l\;T_{b..l..}^{c..k..}
+..
+\Gamma_{ad}^c\;T_{c..j..}^{d..k..}
+..
+\omega_a\;^k\;_l\;T_{b..j..}^{c..l..}
+..
\ee
and $\Gamma,\omega$ themselves by $D_a\; q_{bc}=D_a\; e_b^j=0$. 

The Riemann
tensor of $q$ is defined by the usual formula
$[D_a,D_b]u_c=R_{abc}\;^d u_d$ and $R_{ab}=R_{acb}\;^c,\; R=R_{ab}\; q^{ab}$
Ricci tensor and scalar respectively. Then by letting $D_a,D_b$ act 
only on the tensor indices and using the definition of $\omega$ one finds 
the relation with the curvature $\Omega$ of $\omega$
\be \label{2.2}
R_{abc}\;^d\; e_d^j=-\Omega_{ab}\;^j\;_k\; e^k_c,\;\;
\Omega_{abjk}:=2\partial_{[a}\omega_{b]jk}
+\omega_{ajl}\;\omega_b\;^l\;_k 
-\omega_{akl}\;\omega_b\;^l\;_j
\ee
i.e. $R_{abcd} e^c_j\;e^d_k=\Omega_{abjk}$ and $e^a_j=q^{ab}\delta_{jk} e_b^k$. 

Recall that the canonical formulation of GR without fermionic matter 
can be phrased in terms of $q_{ab}$ and its conjugate momentum \cite{15}
\be \label{2.5}
P^{ab}=\sqrt{\det(q)}\; [q^{ac}\; q^{bd}-q^{ab}\;q^{cd}]\;K_{cd}
\ee
where $K_{ab}=K_{ba}$ be the extrinsic curvature of $\sigma$. We have
(we suppress the Newton constant) 
\be \label{2.6}
\{P^{ab}(x),q_{cd}(y)\}=\delta^a_{(c}\;\delta^b_{d)}\;\delta(x,y)
\ee
all others vanishing. These are subject to the constraints 
\be \label{2.7}
C_a=-2\;D_b\; P^b\;_a,\;\;
C=[\det(q)]^{-1/2}\;[q_{ac}\; q_{bd}-\frac{1}{D-1}\; 
q_{ab}\;q_{cd}]\;P^{ab}\; P^{cd}-
[\det(q)]^{1/2}\; R
\ee
plus matter terms which we suppress in what follows as their detailed 
form does not play any role for what follows. We just remark that 
all bosonic matter terms just depend on $q_{ab}$ while the fermionic 
matter terms require the three fields $K_a^j,\;\omega_{ajk},\;e_a^j$
where $K_a^j$ is introduced below.

We consider 
a mixed field $K_a^j$ whose relation with $K_{ab}$ is given by 
\be \label{2.3}
K_{ab}:=K_{(a}^j\; e^k_{b)} \; \delta_{jk}
\ee
The field $K_a^j$ carries $D^2$ d.o.f. while the symmetric 
tensors $q_{ab},K_{ab}$ carry only $D(D+1)/2$ d.o.f. Both $K_{ab}$ and 
$K_a^j$ carry the same number of degrees of freedom when the 
$D(D-1)/2$ {\it Gauss}
constraints 
\be \label{2.4}
G'_{jk}:=
K_{aj} e^a_k-K_{ak} e^a_j
\ee
vanish which are the same in number as is the dimension of SO(D). Namely 
the vanishing of (\ref{2.4}) implies that the tensor $K_a^j e_{bj}$ 
is automatically symmetric and thus qualifies as extrinsic curvature,
provided $q$ and thus $e$ is not degenerate. In other words
$K_a^j=K_{ab}\; e^{bj}$ when (\ref{2.4}) holds. 

We now define a new canonical pair $(P^a_j, e_a^j)$ and a Gauss constraint 
for it such that the symplectic reduction of that phase space is 
eqivalent to that defined by (\ref{2.6}) and (\ref{2.7}). This 
symplectic structure of $(P^a_j,e_a^j)$ is the outcome of the Lagrangian 
analysis sketched above. It is given by 
\be \label{2.9}
\{P^a_j(x),e_b^k(y)\}=\delta^a_b\;\delta^k_j\;\delta(x,y)
\ee
all others vanishing and the Gauss constraint 
\be \label{2.10}  
G_{jk}=2\;e_{a[j} \; P^a_{k]}   
\ee
The task is now to to show that (\ref{2.6}) is the consequence of 
(\ref{2.9}) and the vanishing of (\ref{2.10}). To see this we define  
$P^{ab},q_{ab}$ as functions of $P^a_j,e_a^j$, then compute their 
Posson brackets using (\ref{2.10}) and check that (\ref{2.6}) results 
plus terms that vanish when (\ref{2.10}) vanishes. This has been 
done e.g. in the first reference of \cite{4} and will not be repeated here,
it suffices to say that the relation between the phase space coordinates 
is given by 
\be \label{2.12}
P^{ab}:=\frac{1}{2}\; P^{(a}_j\; e^{b)j},\;q_{ab}:=e_a^j\;e_{bj}
\ee
%
%$P^a_j=2P^{ab} e_{bj}$ modulo Gauss, thus 
%$P^a_j\;\delta e_a^j=2P^{ab} e^j_{(a} \delta_{b)}=P^{ab}\delta q_{ab}
%
In other words, $P^a_j=2\;P^{ab}\; e_{bj}$ when (\ref{2.10}) holds.
From (\ref{2.5}), (\ref{2.3}) and (\ref{2.12}) we infer
\be \label{2.11}
P^a_j=2\;|\det(e)|\;[K^a_j-e^a_j \; e^b_l\; K_b^l]
\ee
which shows that (\ref{2.4}) and (\ref{2.10}) are equivalent 
when $e$ is non degenerate.  
%
% $P^a_j \delta e_a^j=2 |e| [q^{ab} K_{bj}\delta e_a^j-K_b^l e^b_l 
% e^a_j \delta e_a^j]
% = 2K_{bl} |e| [e^a_j e^{bj} \delta e_a^l-e^b_l e^a_j \delta e_a^j]
% = -2K_{bl} |e|[e^{bj} e_a^l \delta e^a_j+e^b_l e^a_j \delta e_a^j]
% = -2K_{bj} |e|[e^{bl} e_a^l \delta e^a_j+e^b_j e^a_l \delta e_a^l]
% = -2K_{bj} |e|[\delta e^b_j+|e|^{-1}\;e^b_j \delta |e|]
% = -2K_{bj} \delta E^{bj},\;E^a_j=|e| e^a_j
%
Once this is verified we may pass from this $(P,e)$ 
formulation to other canonical coordinates by 
canonical transformations. It is not difficult to see that 
also $(-2 K_a^j, E^a_j=|\det(e)| e^a_j)$ is a canonical pair which 
leads to the $(E,K)$ formulation. Likewise 
$(\tilde{p}_{ab}=-K_{ab}/\sqrt{\det(q)},\tilde{Q}^{ab}=\det(q) q^{ab})$ 
is a canonical pair. 
%
%P^{ab}\delta q_{ab}=|q|^{1/2} K_{ab}[q^{ac} q^{bd}-q^{ab} q^cd]\delta q_{cd}
%=-|q|^{-1/2} K_{ab}[|q|\delta q^{ab}+q^{ab}\delta |q|]
% 
The $(E,K),\; (P,e), (P,q),\;(\tilde{Q},\tilde{p})$ 
formulations are easy to construct in any dimension $D$, 
however the so-called connection $(E,A)$ formulation is available only in 
$D=3$ based on the data 
given above. For $D=3$ one has $A_a^j=K_a^j-\epsilon^{jkl}\omega_{akl}/2$
\cite{6} and the Gauss constraint is $(\partial_a E^a +[A_a,E^a])_{jk}$.
However, even for $D\not=3$ one may still construct 
an $(E,A)$ formulation for any $D\ge 2$ 
using SO(D+1) rather than SO(D) \cite{16}. The reason for why this 
is necessary for $D\not=3$ is that a connection is an element of the adjoint 
representation space while the Vielbein is an element of the defining 
representation space and $D(D-1)/2=D$ iff $D=3$. For $D\not=3$,
in order for $(E,A)$ to form a canonical pair one starts by taking 
both to be valued in the representation space of a Lie algebra of dimension 
$N$. This gives $DN$ canonical pairs and the first class
Gauss constraints of that 
Lie group reduces them to $(D-1)N$. Thus we need $(D-1)N-D(D+1)/2$ 
additional constraints as $q_{ab}$ has $D(D+1)/2$ degrees of freedom.
It turns out that $N=D(D+1)/2$ is a natural choice corresponding to 
SO(D+1) and the additional $(D-2)N$ constraints are the so-called simplicity 
constraints $\epsilon^{M_1..M_{D-3} IJKL}   E^a_{IJ} E^b_{KL}$ for $D\ge 3$. 
These are too many but for $D>3$ are redundant. Their solution 
can in all cases be written as 
$E^a_{IJ}=E^a_{[I} \delta_{J]K}\; n^K,\; 
n^K=\epsilon^{I_1..I_D}\epsilon_{a_1..a_D} E^{a_1}_{I_1}..E^{a_D}_{I_D}$
which reduces the number of degrees of freedom to the $D(D+1)$ $E^a_I$ 
and the Gauss constraint $K_a^{[I} E^{aJ]}=0$ 
to the $D(D+1)/2$ pairs $q^{ab}\det(q)=E^a_I E^b_J \delta^{IJ}$
as desired. In that   
case one constructs a hybrid connection $\hat{\omega}_{aIJ}$ such that 
$\hat{D}_a E^b_I=0$ and has $A_a^{IJ}=\hat{\omega}_a^{IJ}+K_a^{[I} n^{J]}$
modulo simplicity constraints and the Gauss constraint
$(\partial_a E^a +[A_a,E^a])_{JK}$.\\
\\ 
To summarise, in all $D\ge 2$ we have five natural cotangential 
bundle polarisations of the phase space:\\
$(E,A),\; (E,K),\;(P,e),\;(P,q),\; \tilde{Q},\tilde{p})$. 
For all $D$ we may choose as rotation group 
SO(D) for the $(K,E)$ and $(P,e)$ polarisation while 
for all $D$ we may choose as rotation group 
SO(D+1) for the $(A,E)$ polarisation. In addition for $D=3$ we may choose 
SO(3) for the $(A,E)$ polarisation. The Gauss constraints are 
respecively $K_{a[j}\;E^a_{k]},\;e_{a[j}\;P^a_{k]},\;
\partial_a E^a+[A_a,E^a]$ where the commutator is with respect to the 
corresponding Lie algebra.

\section{Triangular gauge}
\label{s3}

The triangular gauge is available in any $D$ for the $(K,E)$ and $(P,e)$ 
polarisation and for $D=3$ for the $(E,A)$ polarisation. We first recall some 
basic knowledge from linear algebra and then apply it to manifolds and 
fibre bundles.
\begin{Theorem} \label{th3.1} ~\\
Let $q_{ab}$ be a positive definite matrix on a real $D$ dimensional
vector space. \\
i. There exists a real valued, 
non-singular $D\times D$ matrix $s^j_a$ such that 
\be \label{3.1}
q_{ab}=\delta_{jk} \; s_a^j\; s_b^k
\ee 
ii. Among the solutions of (\ref{3.1}) there are upper and lower 
triangular (so called Cholesky) solutions respectively, that is, 
$s_a^j=0$ for $a>j$ and 
$a<j$ respectively which are unique up to a choice of $D$ signs.\\
iii. The solution of (\ref{3.1}) is unique up to an orthogonal 
transformation $s^j_a\mapsto O^j\;_k\; s_a^k$. 
\end{Theorem}
This result is of course well known \cite{31}. It is still instructive to 
prove it here in elementary terms to have the formulae readily available
as needed below.
\begin{proof}:\\
i.\\
The simplest argument is based on considering the quadratic form 
$q(x,x)=q_{ab}\;x^a\; x^b$ and to perform successive completion of squares
\be \label{3.2}
q(x,x)=q_{11}[x^1+\sum_{a>1} q_{1a} x^a]^2+q'(x',x') 
\ee
where $x'=(x^2,..,x^D)$ and $q'$ is again positive definite otherwise $q$ 
could not be. Set 
\be \label{3.3}
\sum_{a=1}^D\; s^1_a x^a:=\pm \sqrt{q_{11}}[x^1+\sum_{a>1} q_{1a} x^a]
\ee
where we exploited positive definiteness i.e. $q_{xx}>0$.
Iterate with $x'$ and $x^2$ instead of $x^1$ etc. Then 
\be \label{3.4}
q(x,x)=\sum_{j=1}^D\; [\sum_{a=j}^D s^j_a\; x^a]^2=
\delta_{jk} s^j_a\; s^k_b \; x^a \; x^b
\ee
By polarisation and symmetry $q(x,y)=q(y,x)$ 
\be \label{3.5}
q(x,y)=\frac{1}{4}[q(x+y,x+y)-q(x-y,x-y)]
\ee
for all $x,y$. Thus 
\be \label{3.6}
q_{ab}=\delta_{jk} s^j_a\;s^k_b
\ee
As $\det(q)=[\det(s)]^2>0$, $s_a^j$ is regular.\\
ii. The solution (\ref{3.6}) is upper triangular by construction as we 
performed quadratic completion in the order $x^1,..,x^D$ and is 
the only such solution. The only freedom consists in the sign in each 
completion step. Using the 
opposite order yields a lower triangular matrix with the same amount 
of freedom.\\
iii. Inverting the identity 
$\delta_{jk} e^j_a e^k_b=\delta_{jk} f^j_a f^k_b$ by contracting with the 
inverses $f^a_j f^b_k$ with $f^a_j f_a^k=\delta_j^k$ yields
$\delta_{mn} O^m\;_j \;\;O^n\;_k=\delta_{jk}$ i.e. 
$O_j\;^k:=e_a^j \; f^a_k\in O(D)$ is an orthogonal matrix.\\
\end{proof}
\begin{Corollary} \label{col3.1} ~\\
There is a unique upper and lower triangular solution with positive 
diagonal elements.
\end{Corollary}
This is because a regular triangular matrix must have non vanishing 
diagonal elements as its determinant is the product of those. The sign of 
those diagonal matrix elements can be absorbed into an orthogonal matrix.\\
\\
Above statements just need linear algebra. However, on a manifold $\sigma$ and
fibre bundle with gauge (structure) group SO(D) we have to 
specify what we mean by the condition $a\ge j$ as $a$ refers to the 
tangent space structure while $j$ refers to the frame bundle structure and 
thus per se are not related to each other. In other words, while the 
linear algebra construction above can be done in the tangent space 
over each point
of $\sigma$, one could do this in different ways over each point 
of $\sigma$. Also, suppose one has specified the meaning of $j\ge a$ 
for a certain choice of coordinates on $\sigma$, what happens to this 
condition under changes of the coordinates? What we have done 
in the linear algebraic setting above can be transferred 
to the SO(D) bundle as follows: For a choice of  
coordinates $x^a,\;a=1,..,D$ on $\sigma$ and coordinates 
$y^j,\; j=1,..,D$ on $\mathbb{R}^D$
we have the coordinate co-basis $dx^a$ and the Cartesian coordinate basis 
$\partial/\partial y^j$ on $\mathbb{R}^D$ and the Vielbein with respect 
to these coordinate bases has components $e(x)=e_a^j(x) \; dx^a\;
\partial_{y^j}$. For such choice of coordinates the condition
$e_a^j=0,\; a>j$ now has a precise meaning. It also shows that the 
condition is not invariant under changes of coordinates in either 
$\sigma$ or $\mathbb{R}^D$. Accordingly, in order to meaningfully 
impose the triangular gauge, we must specify coordinates 
$x,y$ on $\sigma,\mathbb{R}^D$ respectively wrt which then $e_a^j(x)=0,\;
a>j$ has a precise meaning. 

As $\mathbb{R}^D$ has a global coordinate system, we pick once and for all 
a global Cartesian coordinate system on $\mathbb{R}^D$.
If $\sigma$ also admits a global coordinate system, we pick one of those.
If the differentiable structure of $\sigma$ is not trivial 
we proceed as follows.
\begin{Definition} \label{def3.0} ~\\
Let $(U_I,x_I)_{I\in {\cal I}}$ be an atlas of a manifold  
$\sigma$ with open cover ${\cal U}=(U_I)_{I\in {\cal I}}$. 
Consider the {\it nerve} of ${\cal U}$ \cite{23} i.e. the set of {\it finite} 
subsets ${\cal J}\subset {\cal I}$ such that $U_{{\cal J}}
:=\cap_{J\in {\cal J}} U_J$ is not empty. If the open 
(as only finite intersections 
are involved) $U_{{\cal J}}$ as ${\cal J}$ varies through the nerve
still cover $\sigma$, then this defines the nerve 
refinement of ${\cal U}$. The manifold $\sigma$ is said to be nerve
refinable if it admits a (possibly refined) 
atlas that has a nerve refinement.
\end{Definition}
Nerve refineability reminds of paracompactness (i.e. 
passing to a refinement if necessary 
the cover is locally finite i.e. every point is 
contained in an neighbourhood intersecting only finitely many of the 
$U_I$) but is not obviously implied or equivalent to it. It is 
a kind of compactness notion which to the best of our knowledge 
has not been discussed in the literature. 
\begin{Lemma} \label{la3.0} ~\\
Let $\sigma$ be a nerve refineable manifold. 
Then $\sigma$ admits a partition such that every member of the partition
is contained in a chart. 
\end{Lemma}
\begin{proof}:\\
By assumption, we may pass to the nerve refinement 
of the given open cover $\{U_I\}_{I\in {\cal I}}$. 
This refinement has the property 
that for all $I,J\in {\cal I}$ we have $U_K:=U_I\cap U_J$ also belongs to 
the cover if not empty. Thus by definition of the nerve (otherwise 
we would have to pass to infinite numbers of intersections)
there are ``smallest'' sets $U_I$ with the 
property that for $J\not=I$ either $U_I\cap U_J=U_I$ or 
$U_I\cap U_J=\emptyset$. Collect those mutually disjoint 
``smallest'' sets $U^{(0)}_I:=U_I$ into a set 
$\cal P$, collect the corresponding indices $I$ into $\cal J$ 
and consider for the remaining sets $U_J,\; J\not\in {\cal J}$ 
their complement 
with respect to all the sets in $\cal P$, i.e. the sets 
$U^{(1)}_J=U_J-\cup_{S\in {\cal P}}\;S$. While these are potentially 
no longer open, they cover $\sigma-\cup_{S\in {\cal P}}\; S$ and 
$U^{(1)}_L=U^{(1)}_J\cap U^{(1)}_K=U_J\cap U_K-\cup_{S\in {\cal P}}$ 
is one of those 
complement sets if not empty. Again by definition of the nerve we        
find ``smallest'' sets among the $U^{(1)}_J$ which are thus mutually disjoint 
and disjoint from the $S\in {\cal P}$. We add those to $\cal P$ and
the corresponding indices to $\cal J$ and consider for the remaining 
$K\not\in {\cal J}$ the sets $U^{(2)}_K=U_K-\cap_{S\in {\cal P}}$ and   
iterate. If the cardinality of the sets in the nerve is uniformly bounded
by a number $N$ then the iteration stops after $N$ steps, otherwise 
it involves a countable number of steps. The end result is a cover 
${\cal P}:=\{V_J\}_{I\in {\cal I}}$ of $\sigma$ where $\cal J$ has become 
$\cal I$ and which is a partition of $\sigma$ by the 
$V_I=U^{(k(I))}_I,\;
I\in {\cal I}$ where $k(I)$ is the step in the above iteration at which 
the index $I$ was collected into the set $\cal J$. Accordingly
$V_I\subset U_I$ for any $I\in  {\cal I}$.\\
\end{proof}
\begin{Definition} \label{def3.0a} ~\\
If $\sigma$ admits a nerve refined atlas 
$(x_I,U_I)_{I\in {\cal I}}$ we consider the corresponding partition 
$(V_I)_{I\in {\cal I}}$ constructed in the lemma. Since $V_I\subset U_I$ 
consider the D-metric $q_I$ expressed in the coordinates $x_I$ over 
$U_I$ and define a D-Bein over $V_I$ to be upper triangular if 
$[e_I]_a^j=0,\; a>j$ with respect to that coordinate system $x_I$. 
By theorem \ref{th3.1} there exists a unique 
upper triangular D-bein $u_I$ 
over $V_I$ with positive diagonal entries 
for all $I\in {\cal  I}$ called the upper triangular 
D-bein relative to the nerve refined atlas.  
\end{Definition}
If $e$ is a given D-bein then $q_{Iab}=
e^j_{Ia}\delta_{jk} e^j_{Ib} 
=u^j_{Ia}\delta_{jk} u^j_{Ib}$ over $V_I$. Thus there exists $g_I\in$ O(D)
over $V_I$ such that $e_I=u_I\cdot g_I$. Given $U_I$ we can 
partition it into $V_K\subset U_K\subset U_K\cap U_I$. 
Let 
$\varphi_{KI}=x_I\circ x_K^{-1}:\;x_K(V_K)\mapsto x_I(V_K)$ be the 
corresponding coordinate transformation. Then 
by definition of the frame bundle we find transition 
functions $g_{KI}\in$ O(D) 
over $x_K(V_K)$ such that 
\be \label{3.6a}
e_K(x)\cdot g_{KI}(x)=
u_K(x)\cdot g_K(x)\; g_{KI}(x)=
[\varphi_{KI}^\ast e_I](x)   
\ee
which expresses $e$ in terms of $u$.\\ 
\\
It follows that one can construct a unique 
upper triangular Vielbein with positive diagonal elements 
relative to a nerve refined atlas
without invoking
characteristic polynomials or the spectral theorem just using completion 
of squares. Note that a triangular matrix has $D+1/2D(D-1)=1/2D(D+1)$
degrees of freedom, i.e. as many as a symmetric matrix has. 

Let us illustrate this explicitly for the important case 
$D=3$. We are given the system of six equations for the six unknowns
$e^3_z;\;e^2_y,e^3_y;\; e^1_x,\;e^2_y,\;e^3_z$ with $e^1_y=e^1_z=e^2_z=0$ 
i.e. $e^j_a=0$ for $a>j$ with $a$ labelling rows and $j$ labelling columns
(we write $a,b=x,y,z$ instead of
$1,2,3$ to distinguish from $j,k=1,2,3$)
\ba \label{3.7}
&& q_{zz}=[e^3_z]^2,\; q_{yz}=e^3_y\; e^3_z,\;q_{xz}=e^3_x\; e^3_z
\nonumber\\
&& q_{yy}=[e^2_y]^2+[e^3_y]^2,\; q_{xy}=e^2_x\; e^2_y+e^3_x e^3_y
\nonumber\\
&& q_{xx}=[e^1_x]^2+[e^2_x]^2+[e^3_x]^2
\ea
These can be successively solved for 
\ba \label{3.8}
&& e^3_z=\sqrt{q_{zz}},\;
e^3_y=\frac{q_{yz}}{e^3_z},\; e^3_x=\frac{q_{xz}}{e^3_z},\; 
\nonumber\\
&& e^2_y=\sqrt{q_{yy}-[e^3_y]^2},\;
e^2_x=\frac{q_{xy}-e^3_x\; e^3_y}{e^2_y}
\nonumber\\
&& e^1_x=\sqrt{q_{xx}-[e^2_x]^2+[e^3_x]^2}
\ea
where we obeyed the positivity of the diagonal elements. Let us check that 
the square roots defined are real. Since a matrix is positive definite 
iff every submatrix has positive determinant we have $q_{zz}>0$ and
\ba \label{3.9}
&& q_{yy}-[e^3_y]^2=\frac{q_{yy} q_{zz}-[q_{yz}]^2}{q_{zz}}>0
\nonumber\\
&& 
q_{xx}-[e^2_x]^2-[e^3_x]^2=\frac{\det(q)}{[e^2_y\;e^3_z]^2} >0 
\ea
\begin{Definition} \label{def3.1} ~\\
An upper (lower) triangular Vielbein (relative to 
a nerve refined atlas) with positive diagonal entries 
is said to be the upper (lower) positively diagonal gauge
UTPD (LTPD) gauge. Explicitly $e_a^j=0,\; a>j$ ($e_a^j=0,\; a<j$) 
and $e_a^j>0,\;a=j$.
\end{Definition}
\begin{Corollary} \label{col3.2} ~\\
For a positive definite metric $q$ the UTPD gauge on the Vielbein $e$
is a perfect gauge, that is, it intersects each gauge 
orbit precisely once and can be reached from any point in the gauge orbit.
\end{Corollary} 
The same statement holds true for any of the $2^D$ gauges corresponding
to choosing the Vielbein to be upper or lower triangular and with any 
of its diagonal entries being positive or negative.\\
\begin{proof}:\\   
A gauge orbit is labelled by a metric $q$. Pick any Vielbein $f$ in that 
gauge orbit, i.e. $q_{ab}=\delta_{jk} f^j_a\; f^k_b$. By above theorem,
there exists a unique Vielbein $e$ in the UTPD gauge such that 
$q_{ab}=\delta_{jk}\; e^j_a\; e^k_b$ and $O\in O(D)$ such that 
$f_a^j=O^j\;_k e^k_a$. 
\end{proof}
Note that because $q$ is non-degenerate, then the existence of a Vielbein
$f$ with $q_{ab}=\delta_{jk} f^j_a f^k_b$ implies that $f$ is non-degenerate 
at every point in $\sigma$ and smooth by smoothness of 
$q$. Thus $\wedge^D f$ is a nowhere vanishing D form on $\sigma$,
hence $\sigma$ is orientable. In that case we may restrict to Vielbeins 
with positive sign of $\det(f)$ which also nowhere vanishes and thus 
the matrix $O$ that establishes the UTPD gauge is necessarily in 
SO(D) rather than O(D). Thus in what follows we restrict to 
orientable $\sigma$.

\subsection{Triangular gauge and diffeormorphism covariance}
\label{s3.1}

We have the following sitation: Let $V$ be the space of all non-degenerate 
D-beins $e$, $M$ the space of all positive D-metrics $q$ and $T$ the 
space of all UTPD D-beins $t$. Note that none of these spaces is a 
$C^\infty(\sigma)$ module due to the non-degeneracy and positivity assumption
but $M$ is a $C^\infty_+(\sigma)$ module (linear combinations with 
postive functions as coefficients). Moreover trivially $T\subset V$.

Then we have maps 
\be \label{3.10}
m:\;V\to M;\; e\mapsto q_{ab}=e_a^j\; e_{bj},\;\;   
u:\;M\to T;\; q\mapsto t_a^j;\; q=m(t), \; t^a_j=0\;\forall a>j,\;
t^a_j>0\;\forall\; a=j
\ee
where $u$ has been constructed explicitly above. Obviously we have 
\be \label{3.11}
m\circ u={\rm id}_M
\ee
but 
\be \label{3.11a}
u\circ m\not={\rm id}_V
\ee
However, denoting the restriction of $m$ to $T$ my $m_T$ we have 
\be \label{3.11b}
u\circ m_T={\rm id}_T
\ee

Diffeomorphisms $\varphi\in {\rm Diff}(\sigma)$ act on $V,M$ by the 
pull-back map
\be \label{3.12}
[\varphi^\ast e^j]_a(x)=\varphi^b_{,a}(x)\; e^j_b(\varphi(x)),\;
[\varphi^\ast q]_{ab}(x)=
\varphi^c_{,a}(x)\; \varphi^d_{,a}(x)\; q_{cd}(\varphi(x)),\;
\ee
and this action passes through the map $m$
\be \label{3.13}
m(\varphi^\ast e)=\varphi^\ast m(e)
\ee
However the map $\varphi^\ast$ does not preserve $T$, it maps
$T$ into $V$. Rather, 
the action of diffeomorphisms on $T$ must preserve the UTPD gauge 
and is given by
\be \label{3.14}
\varphi\cdot t=u(m(\varphi^\ast t))=u(\varphi^\ast(m(t)))
\ee
Due to the highly non-linear, even non-polynomial expression 
for $u$ given above, the action (\ref{3.14}) is rather complicated.
Still we have, using (\ref{3.11})
\ba \label{3.15}
&& [\varphi_1\circ \varphi_2] \cdot t
=u(\varphi_2^\ast\circ \varphi_1^\ast m(t)) 
=u(\varphi_2^\ast m(\varphi_1^\ast t)) 
\nonumber\\
&=& u(\varphi_2^\ast m(u(m(\varphi_1^\ast t))) 
=\varphi_2\cdot u(m(\varphi_1^\ast t))
=\varphi_2\cdot \varphi_1 \cdot t
\ea
i.e. the composition of actions is the action of the compositions 
\be \label{3.16}
[\varphi_1\circ \varphi_2]\cdot=
\varphi_2\cdot \varphi_1\cdot
\ee
just as for the pull-back maps.

Since, using again (\ref{3.11}) 
\be \label{3.17}
m(\varphi\cdot t)=m(u(m(\varphi^\ast t)))=m(\varphi^\ast t)
\ee
there is a unique SO(D) transformation $g$ between 
the D-beins $\varphi\cdot t$ and 
$\varphi^\ast t$
\be \label{3.18}
\varphi^\ast t= [\varphi\cdot t]\;g
\ee
which depends on both $t,\varphi$ and not only $\varphi$ as otherwise 
we would have a representation  of differomorphisms by orthogonal 
transformations and thus would yield multi-valued representations thus 
violating Cartan's obstruction theorem. 

The significance of (\ref{3.18}) is as follows: Suppose we 
are given a function $F$ depending on D-beins $e_a^j$, the spin connection 
$\omega_{ajk}$ of $e$, the field $K_a^j$, spinor fields $\psi$ etc. 
which transforms as a scalar (density) under diffeomorphisms and is Gauss 
invariant which is the case for instance for the contribution 
to the Hamiltonian constraint from the Dirac action. We split the 
D-bein uniquely as $e_a^j=g^j\;_k\; t_a^k\;,\; t=u(m(e))\in T,\; 
g\in$ SO(D). Then $t$ is Gauss invariant (because it depends only on 
the metric) while $g$ transforms as 
$g^j\;_k\mapsto O^j\;_l\; g^l\;_k$ under Gauss transformations $O$.
By means of $g$ we may construct the SO(D) invariant quantities 
\be \label{3.19}
q_{ab},\;\hat{\psi}=\pi(\hat{g}^{-1}) \psi,\;
\hat{\omega}=g^{-1}\;dg+g^{-1} \omega g,\; \hat{K}=g^{-1}K
\ee
etc. very similar to the way one constructs isospin 
gauge invariant vector bosons and fermions out of the Higgs field
dublett \cite{21} 
(one constructs from the complex dublett a Higgs field 
dependent SU(2) matrix and an invariant scalar. Here we construct a 
Vielbein dependent SO(D) matrix and an invariant UTPD Vielbein).
Here $\hat{g}$ is a lift of $g$ to the universal cover of SO(D) whose 
existence is secured by the assumed spin structure. 
Then $F$ can be written in tems of the invariants (\ref{3.19}) but 
these have a non-tensorial transformation behaviour under diffeomorphisms.
Yet, the deviation from a tensorial transformation can be absorbed into 
an SO(D) gauge transformation under which $F$ is invariant by assumption,
and thus $F$ still transforms as a scalar (density) although it just depends 
on the D-Metric and other gauge invariants in a complicated way.\\
\\
It is instructive to explore the action $\varphi\cdot$ in 2d where the 
formulae are still simple while the the non-trivial mechanism becomes 
transparent. We write $x,y$ for $a=1,2$. 
Given $t\in T$ i.e. $t_y^1=0$ we define 
\be \label{3.20}
A=[\varphi^\ast t^1]_x,\;
B=[\varphi^\ast t^2]_x,\;
C=[\varphi^\ast t^1]_y,\;
D=[\varphi^\ast t^2]_y,\;
\ee
and given $q$ we have $t=u(q)$ with 
\be \label{3.21}
t^2_y=\sqrt{q_{yy}},\;t^2_x=\frac{q_{xy}}{\sqrt{q_yy}},\;
t^1_x=\sqrt{q_{xx}-\frac{q_{xy}^2}{q_{yy}}}    
\ee
Thus 
\ba \label{3.22}
A' &=& \varphi\cdot t^1_x=\sqrt{[\varphi^\ast q]_{yy}}=\sqrt{C^2+D^2}
\nonumber\\
B' &=& \varphi\cdot t^2_x=
\frac{[\varphi^\ast q]_{xy}}{\sqrt{[\varphi^\ast q]_{yy}}}
=\frac{AC+BD}{\sqrt{C^2+D^2}}
\nonumber\\
D' &=& \sqrt{[\varphi^\ast q]_{xx}-[B']^2}
=\frac{AD-BC}{\sqrt{C^2+D^2}}
\ea
Then the SO(2) matrix $g$ in (\ref{3.18}) 
solves  
\be \label{3.23}
\left( \begin{array}{cc}
A & B\\
C & D
\end{array}
\right)
=
\left( \begin{array}{cc}
A' & B'\\
0 & D'
\end{array}
\right)
\;\;
\left( \begin{array}{cc}
c & -s\\
s & c
\end{array}
\right)
\ee
where $c=\cos(\phi),\; s=\sin(\phi)$ and one may check that the unique 
solution is given by  
\be \label{3.24}
c=\frac{D}{\sqrt{C^2+D^2}},\; 
s=\frac{C}{\sqrt{C^2+D^2}}
\ee
Obviously $g=1_2$ when $C=0$ which is the case when $\varphi^x_{,y}=0$. 
The set of such diffeos does not form a subgroup however. Formula
(\ref{3.24}) elucidates the dependence of the rotation angle on the 
Vielbein.  \\
\\
Remark:\\
In the reduced phase space approach to quantum gravity one solves 
all constraints by fixing a gauge. For instance using material 
reference systems \cite{12} the spatial diffeomorphism constraint 
is solved by usind $D$ scalar fields as coordinates. This 
is not in contradiction to picking an UTPD gauge for the D-Bein.
It just means that the D scalar fields disappear from whole 
description and that spatial diffeomorphisms no longer act as 
gauge symmetries. After having reduced the spatial diffeomorphism 
constraint one still has the manifold $\sigma$ described by an 
atlas and fixed coordinate systems over each chart. The above 
partition construction now simply tells what an UTPD D-bein is 
in terms of those material reference frame coordinates
rather than some otherwise chosen ones.

\section{Symplectic reduction of the Gauss constraint}
\label{s4}

It remains to solve the Gauss constraint, to compute the reduced 
symplectic structure and to express the remaining constraints in the 
reduced canonical coordinates. We will give all details for the 
$(P,e)$ polarisation and then just describe the mild changes necessary
for the $(E,K), \; (E,A),\; (P,q),\; (\tilde{Q},\tilde{p})$ polarisations. 

\subsection{Solving the Gauss constraint}
\label{s4.1}

In the presence of fermionic matter, the Gauss constraint takes the form 
\be \label{4.1}
G_{jk}=e_{a[j}\; P^a_{k]}+J_{jk}
\ee
where $J_{jk}$ is a fermionic current density typically 
consisting of linear 
combinations of terms of the form $(\Psi^T)^\ast\;[\gamma_j,\gamma_k] \Psi$ 
where $\Psi$ denotes a fermionic species and $\gamma^I$ are Dirac matrices
representing the $D+1$ dimensional Clifford algebra. Due to the 
occurrence of fermionic matter we then have to assume that in addition 
to $\mathbb{\sigma}$ being orientable also 
$\mathbb{R}\times \sigma$ carries as spin structure \cite{15} in order 
to be able to lift SO(1,D) to its universal cover.

We may read (\ref{4.1}) as saying that the antisymmetric part of the 
tensor $P_{jk}:=P^a_j\; e^a_k$ is fixed to be $-J_{jk}$ 
while the symmetric part is unconstrained. Thus we solve (\ref{4.1}) 
by 
\be \label{4.2}
P^a_j
= P_{jk} \; e^{ak}
= P_{(jk)} \; e^{ak}+P_{[jk]} \; e^{ak}
= P_{(jk)} \; e^{ak}-J_{jk} \; e^{ak}
\ee
However, this is not very explicit. One would expect that if $e$ is 
in UTDP gauge, i.e. $e_a^j=0$ for $a>j$ then one could solve the 
$D(D-1)/2$ Gauss constraints for the $D(D-1)/2$ momenta $P^a_j,\; a>j$.
This is indeed the case. To see this, note that 
\be \label{4.3}
(D-1)!\;\det(e)\; e^a_j=\epsilon^{ab_1..b_{D-1}}\;
\epsilon_{jk_1..k_{D-1}}\; e_{b_1}^{k_1}..e_{b_{D-1}}^{k_{D-1}}
\ee
defines the inverse Vielbein.
The expression for $e^a_j$ depends on $e_b^k$ with $b\not=a$ and 
thus in particular those $b>a$ if $a<D$. Those $e_b^k$ are constrained 
to $k\ge b$, thus (\ref{4.3}) certainly involves 
$\epsilon_{jk_1..k_a, a+1, .. D}$. Thus for $j>a$ 
(\ref{4.3}) vanishes, i.e. $e^a_j=0$.
One could call $e^a_j$ therefore LTPD but one should not confuse
the following: We {\it call} $e^a_j$ the inverse Vielbein and it satisfies 
$e_a^j \; e^b_j=[e\; (e^{-1})^T]_a\;^b=\delta_a\;^b$ 
Our definition of upper triangularity $e_a^j=0,\;j<a$ identifies $a$ with 
a row index and $j$ with a column index. In the matrix product 
$e\; (e^{-1})^T=1_D$, if we make the same identification of row and column 
indices for $e^a_j$ then $e^{-1}$ is LTPD but $(e^{-1})^T$ is UTPD. Thus 
there is no contradiction to the statement that the inverse of a regular 
upper (lower) triangular matrix is also upper (lower) triangular (in fact 
the upper (lower) triangular matrices form a closed matrix subalgebra 
of $GL(D,\mathbb{R})$ considered as an algebra and 
the non degenerate ones a subgroup of $GL(D,\mathbb{R})$).    

We now solve the Gauss constraint explicitly for the $P^a_j$ with $a>j$.
By definition in the UTPD
\be \label{4.4}
P^a_j\; e_{ak}-P^a_k\; e_{aj}=-J_{jk}
=\sum_{k\ge a}\;P^a_j\; e_{ak}
-\sum_{j\ge a}\; P^a_k\; e_{aj}
\ee
and we can restrict to $j<k$ as the linearly independent 
equations. We isolate the terms involving $P^a_j,\; a>j$
\be \label{4.5}
\sum_{k\ge a>j}\;P^a_j\; e_{ak}
%-\sum_{j\ge a>k}\; P^a_k\; e_{aj}: does not exist for j<k
%=-[J_{jk}+\sum_{k>j\ge a}\;P^a_j\; e_{ak}-\sum_{k>j\ge a}\; P^a_k\; e_{aj}]
=-[J_{jk}+\sum_{k>j\ge a}\;[P^a_j\; e_{ak}-P^a_k\; e_{aj}]]=:-J'_{jk}
\ee
Note that the equations for different $j$ decouple,
so at fixed $j\le D-1$ we just have to solve the equations for 
$k=j+1,..,D$ starting with $k=j+1$. The equation for $k$ 
involves $P^{x_{j+1}}_j,..,P^{x_k}_j$ and the coefficient of 
$P^{x_k}_j$ in that equation given by $e_{x_k}^k\not=0$. Accordingly,
we can successively solve the $k>j$ equation for $P^{x_k}_j$ as the 
$P^{x_{k'}}_j$ with $k>k'>j$ were already determined starting with 
$P^{x_{j+1}}_j=-[e_{x_{j+1}}^{j+1}]^{-1}\;J'_{j,j+1}$.
 
Thus the solution of (\ref{4.5}) exists, is unique and can be worked
out explictly algebraically. Let us illustrate 
this for the important case $D=3$ for which 
we need only $j=1; k=2,3,\;\;j=2; k=3$
\be \label{4.6}
P^z_2 = - \frac{J'_{23}}{e_z^3},\;\;
P^y_1 = -\frac{J'_{12}}{e_y^2},\;\;
P^z_1 = -\frac{J'_{13}-P^y_1 e_y^3}{e_z^3}
= -\frac{J'_{13} e_y^2-J'_{12} e_y^3}{e_y^2\;e_z^3}
\ee
For an alternative way to solve the Gauss constraint, see (\ref{4.7}) 
below.

\subsection{Reduced symplectic structure}
\label{s4.2}

The Gauss reduced phase space is very simple: 
All canonical pairs and their Poisson brackets 
are kept untouched except that the pairs $(P^a_j,e_a^j),\;a>j$ 
are dropped. Equivalently, one may compute the Dirac bracket between 
functions $f,g$ of the pairs 
$(P^a_j,e_a^j),\;a\le j$ only. The Dirac bracket in this 
case differs from the Poisson bracket by terms $\propto 
\{G_{jk},f\}\;\{e_a^l,g\},\;
\{G_{jk},g\}\;\{e_a^l,f\}$ for $a>l$ both of which vanish. 

More in detail, using the decomposition $e_a^j=g^j\;_k\; u_a^k$
with the UTPD D-bein $u$ and the Gauss invariants 
$\hat{P}^a_k=g^j\;_k P^a_j,\; \hat{\psi}=\pi(g^{-1})\psi$ we have 
\ba \label{4.6a}
&& P^a_j \delta e_a^j
=\hat{P}^a_j \delta u_a^j+P^a_j e_{al} \; g^{lk} \delta g^j_k
\nonumber\\
&=&\hat{P}^a_j \delta u_a^j+P^a_{[j} e_{al]} \; \theta^{jl}
\nonumber\\
&& (\psi^\ast)^T\; \delta \psi- (\delta\psi^\ast)^T\; \psi
=(\hat{\psi}^\ast)^T\; \delta \hat{\psi}- (\delta\hat{\psi}^\ast)^T\; 
\hat{\psi}
+(\psi^\ast)^T\; 
[\pi(g^{-1})\delta\pi(g)-(\delta\pi(g^{-1}))\pi(g)]\hat{\psi}
\nonumber\\
&=& (\hat{\psi}^\ast)^T\; \delta \hat{\psi}- (\delta\hat{\psi}^\ast)^T\; 
\hat{\psi}
+J_{jl} \theta^{jl}
\nonumber\\
\theta^{jl} &=&
g^{lk}\delta g^j_k=-g^{jk}\delta g^l_k
\ea
Thus the two terms proportional to the ``momentary angular velocity''
$\theta^{jl}$ combine to the Gauss constraint $G_{jl}=P^a_{[j} e_{al]}
+J_{jl}$. Note that $\hat{P}^a_j, u_a^j,\; a\le j$ are canonical 
transformations of the ADM coordinates $P^{ab}, q_{ab}$ via 
\be \label{4.6b}
q_{ab}=\delta_{jk} u^j_a u^k_b,\; 
P^{ab}=\hat{P}^c_j [\frac{\partial u_c^j}{\partial q_{ab}}]_{q=u^T u}
\ee
It follows that the spatial diffeomorphism constraint can be written 
just as the spatial diffeomorphism constraint in terms of ADM variables 
plus the $\hat{\psi}$ contribution which tells that $\hat{\psi}$ is a 
scalar density of weight $1/2$ plus a term proportional to the Gauss
constraint.

\subsection{Remanining constraints in terms of the reduced coordinates}
\label{s4.3}

The functions $(e_a^j,\;P^a_j),\; a\le j$ (as well as the fermionic d.o.f.)
are themselves not Gauss invariant. We make any function 
of them Gauss invariant using 
the gauge invariant projection \cite{18}
$[f\mapsto O_f=\exp(X_{\hat{G}(\Lambda)})\cdot f]_{\Lambda=-F}$ where 
$F_a^j=e_a^j=0,\;a>j$ is the gauge fixing condition and 
\be \label{4.a}
\hat{G}^a_l=[\Delta^{-1}]^{jka}_l\; G_{jk},\; \Delta_{jka}^l= 
\{G_{jk},e_a^l\};\;j<k,\; a>l
\ee
The $D(D-1)/2 \times D(D-1)/2$ matrix $\Delta$ is non-degenerate 
\be \label{4.b}
\Delta_{jka}^l=2\delta^l_{[j} e_{ak]}
\ee
since for antisymmetric $M^{jk}$ the equation $M^{jk} \Delta_{jka}^l=0$ 
with $a>l$ gives $M^{lk} e_{ak}=0$ for $a>l$ at $e_{ak}=0, a>k$ i.e.
$\sum_{k\ge a>l} M^{lk} \; e_{ak}=0$ which can be solved uniquely for 
$M^{lk}=0$. We used 
$\hat{G}(\Lambda)=\int\; d^Dx\;\Lambda_a^j\; \hat{G}^a_j$ and $X_H$ denotes 
the Hamiltonian vector field of $H$.   

Using $H=H(N,\xi)=C(N)+C_a(\xi^a)$, i.e. the linear combination of constraints other 
than the Gauss constraint we have 
\be \label{4.c}
\{O_H,O_f\}\approx \{H,O_f\} \approx O_{\{H,f\}^\ast}
\ee
where $\{.,.\}$ is the Dirac bracket of $G_{jk},\; F_a^l$ and $\approx$ means 
(weak) 
equality modulo linear combinations of the Gauss constraint. We have used that 
$H$ is itself weakly Gauss invariant. Splitting 
the degrees of freedom as 
$x=(e_a^j)_{a>j},\; r=(e_a^j)_{j\ge a},\;
y=(P^a_j)_{a>j},\; s=(P^a_j)_{j\ge a}$, one has 
for any function $g=g(x,y,r,s)$ that 
$O_g=g(0,L(0,O_r,O_s),O_r,O_s)$ where $y=L(x,r,s)$ is the solution of 
$G(x,y,r,s)=0$ constructed in the previous subsection. Accordingly
the algebra of observables is coordinatise by $O_r,O_s$ and 
$\{O_r,O_s\}=O_{\{r,s\}^\ast}=O_{\{r,s\}}=\{r,s\}$ as $\{F,r\}=\{F,s\}=0$
and $\{r,s\}$ is a constant on phase space. 

Thus, abusing the notation and writing $r,s$ instead of $O_r,O_s$ and 
$g(r,s)$ instead of $O_g=g(O_r,O_s)$ we find that we can simply 
use the constraints $C,C_a$ as before but with the substitution
of $x,y,r,s$ by $x=0, y=L(0,O_r,O_s),O_r,O_s)$ or $(0,L(0,r,s),r,s)$ 
under this abuse of notation where $r,s$ still have canonical brackets. 
We also can still use the Poisson bracket rather than the Dirac bracket.
Since $\{H(N,\xi),H(N',\xi')\}$ is a linear combination of such $H$ again
the same is true for $\{O_H,O_{H'}\}=O_{\{H,H'\}^\ast}\approx O_{\{H,H'\}}$.

To summarise we may simply insert the solution of the Gauss constraint and 
the UTPD gauge fixing condition into spatial diffeomorphism and Hamiltonian 
constraint and use the Poisson bracket on the unconstrained variables.
We remark that the Hamiltonian constraint for the bosonic sector can 
be taken as the ADM expression and for the fermionic sector is given by 
the Dirac Hamiltonian which is manifestly Gauss invariant. 
For the spatial diffeomorphism constraint one 
has $C_a(\xi^a)=\int\;d^Dx\; P^a_j\; [{\cal L}_\xi e^j]_a$ where 
${\cal L}_\xi$ is the Lie derivative which shows that it has weakly vanishing
brackets with the Gauss constraint (which is a scalar density of weight one).
We also remark that all this can be alternatively seen as follows: The 
above discussion reveals that one should use the Dirac bracket on the 
gauge fixed phase space. On functions $g$ of $r,s$ the Dirac bracket coincides
with the Poisson bracket. The Dirac bracket between $H$ and a function 
of $r,s$ does not reduce to the Poisson bracket but receives the 
correction schematically given by 
\be \label{4.c}
\{H,g\}^\ast-\{H,g\}=
\{H,G\}\;\cdot\Delta^{-1}\cdot\{F,g\}-  
\{g,G\}\;\cdot\Delta^{-1}\cdot\{F,H\}=-
\{G,g\}\;\cdot\Delta^{-1}\cdot\{H,x\}
\ee
which at $x=0,y=L$ coincides with (use $G=\Delta\cdot (y-L),\;\{g,y\}=0$
\be \label{4.d})
\{H_{x=0,y=L},g\}
=\{H,g\}_{x=0,y=L}+\{H,x\}_{x=0,y=L}\cdot\;\{L,g\}  
=\{H,g\}_{x=0,y=L}-\{H,x\}_{x=0,y=L}\cdot\;\Delta^{-1}\;\cdot\{G,g\}  
\ee    
i.e. by substituting $x=0,y=L$ into the Hamiltonian. Moreover
$H=H_{x=0,y=L}+a\cdot F+b\cdot G$ for certain $a,b$ 
thus modulo terms proportional to $x=F, G$ 
\be \label{4.4e}
\{H,H'\}^\ast=\{H_{x=0,y=L},H'_{x=0,y=L}\}^\ast=  
\{H_{x=0,y=L},H'_{x=0,y=L}\}
\ee
as $H_{x=0,y=L},H'_{x=0,y=L}$ only depend on $r,s$.\\
\\
\\
The strategy to write the constraints $C_a,C$ in terms of the reduced 
phase space coordinates is rather simple:\\
1.\\ 
Whereever $e_a^j$ appears, either directly or indirectly as 
$q_{ab}=e_a^j \; e_{bj},\; e^a_j,\;\omega_{ajk}(e)$  etc.
one just restricts it to the UTPD gauge 
$e_a^j=0$ for $j<a$.\\
2.\\ 
Whereever $P^a_j$ appears, either directly or indirectly as 
$P^{ab}=P^{(a}_j e^{b)j},\; 
K_a^j=|\det(e)|^{-1}q_{ab}[P^b_j-e^a_j e_b^k P^b_k]$ etc one 
keeps $P^a_j$ as it is for $j\ge a$ and 
makes for $a>j$ use of the identity 
\ba \label{4.7}
P^a_j
&=&\sum_k\; [P^b_j\; e_b^k]\;e^a_k,\;
=\{\sum_{j\ge k}\; [P^b_j\; e_{bk}]+\sum_{j<k}\;[P^b_j\; e_{bk}]\}\;e^{ak}
\nonumber\\
&=&
\{\sum_{j\ge k}\; [P^b_j\; e_b^k]+\sum_{k<j}\;[-J_{jk}+P^b_k\; e_{bj}]\}
\;e^{ak}
\ea
where the Gauss constraint has been used. Now for $j\ge k$ we have 
\be \label{4.8}
P^b_j e_{bk}=\sum_{k\ge b} P^b_j\; e_{bk}
\ee
i.e. only $P^b_j$ with $j\ge b$ occurs. Thus both sums on the r.h.s. of 
(\ref{4.7}) only depend only on the unconstrained momenta and thus 
yields an alternative way to solve the Gauss constraint explicitly.

\subsection{Other polarisations}
\label{s4.4}

What was said in the previous subsections for the $(P,e)$ polarisation 
almost literally applies also to the other polarisations 
$(E,K),\; (E,A),\;(P,q),\;(\tilde{Q},\tilde{p})$. 
For all polarisations we use the UTPD gauge 
$e_a^j=0$ for $j<a$ which implies that $e^a_j=0$ for $a<j$ and thus 
also $E^a_j=0$ for $a<j$. \\
\\
The Gauss constraint for the 
$(E,K)$ polarisation is given for any dimension $D$ by 
\be \label{4.9}
G_{jk}=2\; K_{[aj}\; E^a_{k]}+J_{jk}
\ee
which maybe explicitly checked using $P^a_j=|\det(e)|[K^a_j-e^a_j\; e^b_k\;
K_b^k]$ and $E^a_j=|\det(e)|\; e^a_j$.\\
\\
The Gauss constraint for the 
$(E,A)$ polarisation is given for $D=3$ by 
\be \label{4.10}
G_{jk}=\partial_a[E^a_l+\epsilon_{lmn} A_a^m\; E^a_n]\epsilon_{jlk}+J_{jk}
\ee
with $A_a^j=K_a^j+\frac{1}{2}\epsilon_{kjl}\;\omega_a\;^{kl}$ as may be 
checked using $D_a E^a_j=\partial_a E^a_j+\omega_{ajk}\; E^{ak}=0$ 
which in turn follows from (\ref{2.1}). \\
\\
For $D>3$ we need the SO(D+1) formulation in the $(A,E)$ polarisation and 
then the UTPD gauge is not sufficient as we need to gauge fix both 
the $D(D+1)/2$ SO(D+1) Gauss constraints and the 
$(D-2)D(D+1)/2$ simplicity constraints using $(D-1)D(D+1)/2$ gauge fixing 
conditions on the $D^2(D+1)/2$ variables $E^a_{IJ},\; 0\le I<J\le D$. 
While this can be done, a UTPD gauge is not sufficient in this case 
and we will not go into further details here.\\
\\
For any $D$ in the $(P,q),\; (\tilde{Q},\tilde{p})$ polarisations, 
we use the Gauss constraint 
of the $(P,e)$ formulation discussed in the previous subsections. Note 
that in the $(P,q)$ polarisation nothing needs to be done at all 
if there is no fermionic matter.\\
\\
The constraints (\ref{4.9}) and (\ref{4.10}) can be solved by exactly 
the same method as in the case of the $(P,e)$ polarisation, just 
that instead of $P^a_j$ for $a>j$ we solve for $K_a^j,\;A_a^j$ for 
$a<j$ respectively and leave $K_a^j,\; A_a^j$ for $a\ge j$ unconstrained.
This can be done most efficiently using 
the analog of (\ref{4.7}) and (\ref{4.8}), i.e. we 
write for $a<j$ (with the inverse $E^a_j E^j_b=\delta^a_b$)
\be \label{4.11}
K_{aj}=\sum_k\; [K_{bj} E^b_k]\; E_a^k
=\sum_{j\le k} [K_b^j E^b_k]\; E_a^k
+\sum_{j> k} [K_b^k E^b_j-J_{jk}]\; E_a^k
\ee
and 
\be \label{4.12}
A_{aj}=\sum_k\; [A_{bj} E^b_k]\; E_a^k
=\sum_{j\le k} [A_b^j E^b_k]\; E_a^k
+\sum_{j> k} [A_b^k E^b_j-(\partial_a E^a_l+ \epsilon_{jlk} \;J_{jk}]\; E_a^k
\ee
where the r.h.s of (\ref{4.11}) and (\ref{4.12}) respectively only depend 
on $K_a^j,A_a^j$ respectively with $a\ge j$. It should be noted 
that solving the Gauss constraint as in (\ref{4.12}) introduces 
derivatives of $E^a_j$ into the reduced Hamiltonian constraint.\\
\\
The respective reduced symplectic structures only depend on the canonical pairs 
$(K_a^j, E^a_j)$,   
$(A_a^j, E^a_j)$ with $a\ge j$ and on all $(P^{ab},q_{ab}),\; 
(\tilde{Q}^{ab},\tilde{p}_{ab})$ 
(as $P^{ab},q_{ab}$ are symmetric tensors) respectively.\\   
\\
The reduced spatial diffeomorphism and Hamiltonian constraints are simply 
given by substituting gauge conditions and solutions of the Gauss constraints.

\section{New representations of the canonical commutation relations}
\label{s5}

\subsection{Motivation and preliminaries}
\label{s5.1}

To motivate the development of new representations of the canonical 
commutation relations, let us review how the concrete represention 
currently used in LQG was incarnated:\\
For a long time canonical QG was studied in the $(P,q)$ polarisation
\cite{19} 
but no rigorous Hilbert space representation of the CCR was found, 
the complicated algebraic form of the Hamiltonian constraint in that 
polarisation in its density weight one form 
was believed a major road block and the $(P,q)$ polarisation 
by itself is unsufficient
to cope with fermionic matter. The new canonical connection variables 
found for $D=3$ in \cite{6} made it possible to simplify the algebraic form 
into a polynomial in its density weight two form and to accomodate 
fermionic matter, however only 
when the connection is complex valued. The associated reality conditions
encapsulate the complicated algebraic form of the $(P,q)$ polarisation 
and so far no representation of the CCR (including the $^\ast$ relations)
could be found (see \cite{11,20} for some attempts). 
In the real valued form \cite{6} and polarisation $(A,E)$ which 
is restricted also to $D=3$ one could find CCR representations \cite{5}
but in its density one form the constraint stays complicated and 
it becomes polynomial only at much higher density weight. In \cite{11}
it was shown that density weight one is preferred and that the complicated 
algebraic form is not a nuisance but a necessity in order to 
define the Hamiltonian densely in the representation \cite{5}. That is,
there are strong dynamical reasons for why one considers connections 
$A$ integrated along 1D paths and electric fields $E$ along 2D 
surfaces. Furthermore,
the reason why one uses the $(A,E)$ plolarisation rather than say the 
$(K,E)$ polarisation is that in the representation \cite{5} using connections 
it is quite straightforward to solve the Gauss constraint in the 
quantum theory using methods from lattice gauge theory while this 
is not the case in the $(K,E)$ polarisation.\\
\\
Accordingly, one of the strongest arguments in support of
using the $(A,E)$ polarisation and non-Abelian 
connections and electric fluxes to formulate the CCR in LQG is 
to have gauge covariant or invariant {\it quantum formulation
of the Gauss consttraint}.
It is thus a natural question which other possibilities arise if 
one {\it solves the Gauss constraint classically}. Given the 
$(\tilde{Q},\;\tilde{p}),\;(P,q),\; (P,e), \; (K,E)$ 
polarisations discussed above, after gauge 
fixing the Gauss constraint, the following appears to be conceivable:\\ 
1.\\
The non-Abelian character of the CCR can  
be dropped thereby avoiding many 
non-commutativity commplications \cite{9}.\\ 
2.\\
Instead of smearing the fields 
with ``sharp support'' along 1D paths and 2D surfaces, it may be possible 
to smoothen the support and thus obtain improved continuity properties 
of the corresponding ``Weyl'' operators.\\
3.\\ Finally, the restriction to 3D maybe avoidable.\\
\\
In what follows we will show that all three possibilities can be realised
with respect to all these four polarisations. One may wonder how 
the $(\tilde{Q},\tilde{q}),\;(P,q)$ polarisations 
can persist in presence of fermionic matter.
The answer is that the ADM phase space coordinatised by 
$(P,q)$ is indeed augmented in the fermionic sector by fields 
$K_a^j, e_a^j, \omega_{ajk}$ with the Gauss constraint still
present but that these can be expressed purely 
in terms of $P^{ab},q_{ab}$ or 
$\tilde{P}^{ab},\tilde{q}_{ab}$ 
when the UTPD gauge is installed and Gauss 
constraint is solved. While this leads to non-polynomial expressions in 
the fermionic sector, see e.g. (\ref{3.7})
these are not worse than those already present in the 
bosonic sector due to (negative) powers of $\sqrt{\det(q)}$ so that 
this complication does not disfavour these polarisations.
In fact, the ADM like polarisations $(P,q),(\tilde{Q},\tilde{p})$ 
are particularly attractive 
because these variables are manifestly Gauss invariant, have a transparent 
spacetime interpretation and are the only polarisations in the list 
above in which 
the bosonic sector does not suffer from major algebraic modification
after installing the UTPD gauge. Finally, it was believed for a long time 
that the ADM like polarisations cannot be used to define rigorous 
Hilbert space representation of the CCR which are also 
useful for the implementation of the Hamiltonian constraint because 
one cannot naturally integrate symmetric tensor (densities) over 
submanifolds. While this is true, by item 2. above one circumvent 
this objection by using integrals over submanifolds not of 
those tensor fields themselves (sharp support) but contracted with 
suitable test tensor fields.

Finally we note that e.g. a 1d integral over a Vielbein in UTPD gauge 
does not behave in the standard way under diffeomorphisms because 
diffeomorphisms act on UTPD D-beine not by pull-back but differently, 
see section \ref{s3.1}. In that respect, representations based on such 
obejects and the Weyl elements that one constructs from them are 
more useful in situations in which one performs a complete gauge fixing 
of all constraints \cite{12} 
including the spatial diffeomorphism and 
Hamiltonian constraints 
or at least also the spatial diffeomorphism
constraints \cite{13a}
so that diffeomorphisms no longer act. On the other hand, note that 
positive definite metrics do not even form a vectctor space therefore 
they seem to be ruled out as a configuration degree of freedom part
in a Heisenberg algebra like approach to quantisation where the Heisenberg
algebra is necessarily a Lie algebra. That objection can be met by 
considering all possible D-metrics as the kinematical configuration space 
(i.e. any signature, including degenerate) and to insist that semiclassical 
states be supported on positive definite metrics. Also, positive definite
metrics may play an important role concerning domain questions of the 
quantum hypersurface deformation algebra \cite{13}. In that respect D-beins 
in the upper diaagonal gauge perform slightly 
better in that they do form a vector space but again the issue non-degeneracy 
must be given up in the kinematical quantisation step and its discussion 
be postponed towards semiclassical and domain questions.  

Before we start describing these alternative polarisations, note that 
there is an asymmetry in the polarisations we listed so far:
While in the $P,e$ and $E,K$ and $P,q$ polarisation the momentum is a 
density of weight one, in the $\tilde{Q},\tilde{p}$ polarisation 
the momentum is a density of weight two which will be inconvenient for 
what follows.
However, it is a simple matter of a canonical 
transformation to introduce yet another $Q,p$ polarisation 
where $Q=\sqrt{\det(q)} q^{ab}$ is a symmetric contravariant tensor 
density of weight one and $p$ a symmetric covariant tensor. We have 
up to total differentials on phase space
\be \label{5.1}
Q^{ab} \delta p_{ab}
=-p_{ab} \delta Q^{ab}
=-p_{ab}\sqrt{\det(q)}[-q^{ac} q^{bd}\; \delta q_{cd}+\frac{1}{2}
q^{ab}\; q^{cd}]\; \delta q_{cd}
\ee
from which 
\be \label{5.2}
P^{ab}=\sqrt{\det(q)}[q^{ac}\;q^{bd}-\frac{1}{2} q^ab\; q^{cd}] \; p_{cd}
\;\;\Leftrightarrow\;\;
p_{ab}=\frac{1}{\sqrt{\det(q)}}[q_{ac}\; q_{bd}-\frac{1}{D-2} q_{ab} q_{cd}]
P^{cd}
\ee

\subsection{Polarisations and Poisson algebra of classical smeared variables}
\label{s5.1}

Among the potentially interesting representations for quantum gravity 
are those for which the metric or Vielbein operator annihilates the vacuum.
This is beccause these operators appear in every single term of the 
Hamiltonian constraint and thus have good chances to turn the Hamiltonian
constraint operator into a densely defined operator on the dense domain 
generated 
by acting on that vacuum by suitable Weyl operators. This means 
that the vacuum $\Omega$ satisfies $e\Omega=0,\; E\Omega=0,\;q\Omega=0,\;
Q\Omega=0$ respectively. Furthermore, as a minimal requirement, we want 
that both the cosmological term 
\be \label{5.3}
\Lambda \int_\sigma\; d^Dx\;N\; \sqrt{\det(q)}
\ee
and the kinetic term 
\be \label{5.4}
\int \;d^Dx\;N\; 
\frac{1}{\sqrt{det(q)}}[q_{ac}q_{bd}-\frac{1}{D-1} q_{ab} q_{cd}]\;P^{ab} P^{cd}
\ee
with lapse test function $N$ 
are separately well defined, at least with an intermediate regularisation. 
Of course both (\ref{5.3}) have to be expressed in terms of the variables 
that the polarisation is based on. We have $\sqrt{\det(q)}=|\det(e)|$ 
and $|\det(E)|=|\det(e)|^{D-1}$ while $\det(Q)=\sqrt{\det(q)}^{D-2}$
i.e. $\sqrt{\det(q)}=|\det(E)|^{1/(D-1)}=\det(Q)^{1/(D-2)}$. In a Riemann 
sum approximation of the cosmological term by cells of coordinate 
volume $\epsilon^D$ we see that 
\be \label{5.5}
\epsilon^D\;\sqrt{\det(q)}
=\sqrt{\det(\epsilon^2\; q)}     
=|\det(\epsilon\; e)\     
=|\det(\epsilon^{D-1} E)|^{1/(D-1)}     
=|\det(\epsilon^{D-2} Q)|^{1/(D-2)}     
\ee
which fixes $q,e,E,Q$ respectively 
to be smeared in $2,1,D-1,D-2$ dimensions respectively.
Next we have (it suffices to consider the respective $q_{ab} P^{ab}$ 
contribution) 
\ba \label{5.6}
&& \; \epsilon^D\; \frac{1}{\sqrt{\det(q)}}\; [q_{ab} P^{ab}]^2
=\frac{1}{\epsilon^D\;\sqrt{\det(q)}}\; [(\epsilon^2\;q_{ab})\;
(\epsilon^{D-2}\;P^{ab})]^2
\nonumber\\
&=& \frac{1}{\epsilon^D\;\sqrt{\det(q)}}\; [(\epsilon\;e_a^j)\;
(\epsilon^{D-1}\;P^a_j)]^2
\nonumber\\
&=& \frac{1}{\epsilon^D\;\sqrt{\det(q)}}\; [(\epsilon^2\;p_{ab})\;
(\epsilon^{D-2}\;Q^{ab})]^2
\nonumber\\
&=&\frac{1}{\epsilon^D\;\sqrt{\det(q)}}\; [(\epsilon\;K_a^j)\;
(\epsilon^{D-1}\;E^a_j)]^2
\ea
which establishes that the respective conjugate variables 
$P^{ab},P^a_j, K_a^j, p_{ab}$ have to be smeared in the respective dual 
dimension $D-2,D-1,1,2$.

Thus, purely dynamical considerations dictate that the smearing 
dimensions of the conjugate veariables add to the dimension of 
the spatial manifold. Now $e,K$ are 1-forms and $\ast P, \ast E$
are (pseudo) (D-1)-forms where e.g. 
$[\ast P_j]_{a_1..a_{D-1}}=P^b_j\; \epsilon_{b a_1.. a_{D-1}}$. 
Therefore these can be, after contraction of the open j index with 
a test function $f^j$, naturally be integrated over 1-submanifolds $C$ and 
(D-1)-submanifolds $S$ respectively
\ba \label{5.7}
<e,f>_S &:=& =\sum_{j\ge a}\; \int_C\; dx^a\;f_j \;e^j_a,\;\;\;
<K,f>_S := \sum_{j\ge a}\; \int_C\; dx^a\;f_j \;K^j_a
\nonumber\\
<P,g>_S &:=& \sum_{j\ge a}\;\int_S\; dx_a\; g^j P_j^a,\;\;\;
<E,g>_S := \sum_{j\ge a}\;\int_S\; dx_a\; g^j E_j^a
\nonumber\\
dx_a &:=& \frac{1}{(D-1)!}\epsilon_{ab_1..b_{D-1}}\; dx^{b_1}\wedge
.. \wedge dx^{b_{D-1}}
\ea
Note that the contraction with $f^j$ 
is not done in LQG for $A$ because there one focuses on holonomies 
for reasons of 
gauge covariance and these 
are constructed with ``sharp support'' on 1-submanifolds. In our context
when the Gauss constraint is already removed, such additional 
smoothened out support enforced by $f^j$ does become available.  

However, the symmetric tensor fields $q_{ab},\;p_{ab},\;P^{ab},\; Q^{ab}$ 
are not 2-forms or dual to (D-2)-forms. Similar to contracting 
$e_a^j, K_{aj}, P^a_j, E^a_j$ with test functions $f_j,f^j$ respectively,
in this case we proceed as follows: Consider test tensors of type 
(1,1) i.e. $f_a^b,\;g_a^b$ and consider for 2-submanifolds $C$ and 
$(D-2)$ submanifolds $S$ 
\ba \label{5.8}
<q,f>_C &:=& \int_S\; dx^a\wedge dx^b f_{[a}^c\; q_{b]c},\;\;\; 
<p,f>_C := \int_S\; dx^a\wedge dx^b f_{[a}^c\; p_{b]c},\;\;\; 
\nonumber\\
<P,g>_S &:=& \int_c\; dx_{ab}\; g^{[a}_c\; P^{b]c},\;\;\;
<Q,g>_S := \int_c\; dx_{ab}\; g^{[a}_c\; Q^{b]c},\;\;\;
\nonumber\\
dx_{ab} &:=& \frac{1}{(D-2)!}\epsilon_{ab c_1..c_{D-2}}\; dx^{c_1}\wedge
.. \wedge dx^{c_{D-2}}
\ea
In (\ref{5.8}) all indices have full index range while in 
(\ref{5.7}) we have the constraint $a\le j$.

To see that the Poisson bracket algebra generated from 
(\ref{5.7}) and (\ref{5.8}) is well defined 
for general 1-, 2-, (D-2)-, (D-1)- manifolds respectively requires  
some extra care as compared to the LQG situation. As in LQG we require 
that all submanifolds are piecewise analytic (or semi-analytic \cite{5}). 
Then for a 1-submanifold
$C$ we partition it into finitely many segments which either lie entirely
inside or outside the $(D-1)-$submanifold $S$ or on the up or down side of 
$S$ relative to the co-normal of $S$. The Poisson bracket between e.g.
$<P,g>_S$ and $<e,f>_C$ can be written as a sum over segment $s$ 
contributions and such a contribution contains an integral of the form 
\be \label{5.9}
\int_0^1\; dt\;\int_{[0,1]^{D-1}}\; d^{D-1}u \; F(t,u)\; 
\delta(s(t),S(u))
\ee
where 
$t\in [0,1]\mapsto s(t)\in \sigma,\; u\in [0,1]^{D-1}\mapsto S(u)\in 
\sigma$ define the embedded 1- and (D-1)-submanifolds. Clearly 
contributions from outside segments vanish. For inside segments 
(\ref{5.9}) is ill-defined because $s(t)=S(U(t))$ for some 
$t\mapsto U(t)\in [0,1]^D$ so that (\ref{5.9}) contains $\delta(0,0)$ after 
performing the $u$ integral. In LQG the function $F(t,u)$ contains 
a factor $n_a^S(u)\; \dot{s}^a(t)$ which vanishes at $u=U(t)$, hence 
there the expression is still ill-defined but of the form $0\cdot \infty$
while 
for (\ref{5.7}) this is not the case due to the $j\ge a$ restriction in 
the summation. Similar remarks hold for (\ref{5.8}). Next, for the 
up or down type segments, it may happen that the tangent of $c$ at 
the intersection point $p$ lies in the tangent space of $S$ at $p$.
In that case solving the $\delta$ distribution produces a 
Jacobean $|\det[\partial(s(t)-S(u))/\partial (t,u)]|_p^{-1}$ which diverges.
In LQG that pole is accompanied by the zero of same order from the 
factor $n_a^S(u)\; \dot{s}^a(t)$ in $F(t,u)$ so that the integral is well 
defined using de l'Hospital's theorem. 
However, again due to the restriction $j\ge a$ that factor 
is missing form (\ref{5.7}) and similar remarks hold for (\ref{5.8}).

To give meaning to the a priori ill-defined (\ref{5.9}) we proceed as in LQG
and regulate it but we use a more sophisticated regularisation:
For the inside segments we consider instead of (\ref{5.9})
\be \label{5.9a}
\lim_{\epsilon\to 0}\;
\int_0^1\; dt\;\int_{[0,1]^{D-1}}\; d^{D-1}u \; F(t,u)\; 
\delta(s(t),S_{\epsilon^{3/2}}(u))=0
\ee
where $\delta\mapsto S_\delta$ is a foliation of surfaces with 
$S_0=S$. Hence $S_\delta\cap s=\emptyset$ for $\delta\not=0$, thus 
(\ref{5.9a}) follows immediately. The justification for (\ref{5.9a}) 
is that the classical $<P,g>_{S_\delta}$ is continuous in $\delta$.
The reason for why we use $\delta=\epsilon^{3/2}$ will become clear
immediately. For the up and down segments respectively 
we replace (\ref{5.9}) by 
\be \label{5.9b}
\lim_{\epsilon\to 0+}\;
\int_0^\epsilon\; dt\;\int_{[0,1]^{D-1}}\; d^{D-1}u \; F(t,u)\; 
\delta(s(t),S_{\pm \epsilon^{3/2}}(u))
\ee
where we used coordinates such that the intersection point $p$ is given by 
$u=0$ and $t=0$.
The justification for (\ref{5.9b}) is again continuity with respect 
to the choice of $S_{\pm \epsilon^{3/2}}$ and that 
we get a contribution to the integral only from an arbitrarily small
neighbourhood of $0$ for $t$, or in other words, the 
segment $s$ of up or down type can be chosen arbitrarily short i.e. 
we may replace it by $s_\delta(t)=s(\delta t),\; \delta>0,\; t\in [0,1]$ 
and we have 
chosen the coincidence limit $\delta=\epsilon$ as part of the regularisation. 
Now in suitable coordinates we have 
$S_{\pm\epsilon^{3/2}}(u)=(u,\pm\epsilon^{3/2})$ and 
$s(t)=(U(t),f(t))$ where $U(0)=0$, $f(t)$ is real analytic in $t$
and we have $f(t)=\pm \kappa t^n+O(t^{n+1})$ with $\kappa>0$ for 
up and down type and $n\in \mathbb{N}$ is the lowest contributing 
order. Formally we can also capture the inside type using $n=\infty$.
The equation $\epsilon^{3/2}=\kappa t^n$ is solved by 
$t=[\epsilon^{3/2}]^{1/n}$ which lies in $[0,\epsilon]$ iff
%$[\epsilon^{3/2}/\kappa]^{1/n}\le \epsilon$
$\epsilon^{3/2}\le \kappa \epsilon^n$. For $n=1$ this means 
$\epsilon^{1/2}\le \kappa$ which is eventually the case for 
all $\kappa^2\ge \epsilon\ge 0$ no matter how small $\kappa$ is. 
For $n\ge 2$ this is eventually not possible for 
$\epsilon^{n-3/2}\le 1/\kappa$ no matter how large $\kappa$ is. 

As a result of this regularisation, (\ref{5.9}) is 
{\it defined to vanish for non-transversal} intersections, i.e. 
for all inside type segments and all up or down segments such that 
at the intersection point $p$ the tangent space of of $S$ together with 
the tangent of $c$ does not span the tangent space of $\sigma$ at $p$. 
We will refer to non-transversal intersections as co-planar.
This can be captured by the indicator number 
\be \label{5.10}
\sigma_p(S,C)=|{\rm sgn}(\det[\partial (C(t)-S(u))/\partial(t,u)])_p|  
\ee
which equals unity for transversal isolated intersections and zero 
for co-planar ones. It extends naturally by zero 
to intersections of inside type.
Note that (\ref{5.10}) is diffeomorphism invariant 
because under diffeomorphisms the determinant just changes by the 
Jacobean of the diffeomorphism which is always positive. We will 
use the same rule for the Poisson brackets of the variables (\ref{5.8}).

With this understanding, the Poisson brackets of (\ref{5.7}) and 
(\ref{5.8}) are well defined. As opposed to LQG,
they will in general contain a 
factor $|\det[\partial (C(t)-S(u))/\partial(t,u)])_p|^{-1}$ 
and depend in addition on the tensor contractions of the smearing 
functions $f^a_j,g_a^j,f_a^b, g^a_b$ with tangent and co-tagent 
directions of the submanifolds that one integrates over.
    
The smearing of symmetric tensor fields ($D(D+1)/2$ independent components)
by mixed type tensor fields ($D^2$ independent components) is slightly 
redundant. However, there appears to be no covariant {\it linear} algebraic 
restriction on such tensors. While we might impose that $f$ should
have $D(D-1)/2$ first vanishing moments 
${\rm Tr}(f^k)=f_{a_1}\;^{a_2}\; ..\; f_{a_k}\;^{a_1},\; k=1,..,D(D-1)/2$
which does not require a background metric, this condition is 
not linear unless $D=2$ and thus the space of these smearing functions 
would not be a vector space preventing the exponentials of 
integrals of (\ref{5.7}), (\ref{5.8}) to form a Weyl algebra. Thus 
we keep the mixed tensors unconstrained, noting that they suffice 
to separate the points of the phase space. To see this, fix some 
Vielbein $v_a^j$ and note that we may pick $S$ such that 
$dx^a\wedge dx^b$ be along $2\; v^{[a}_i v^{b]}_j$ and 
$f_a^c$ along $v_a^k \; v^c_l$ for any choice of $i<j$ and $k,l$ so that 
(\ref{5.8}) allows to extract $2 \delta^k_{[i} q_{j]l}$ and picking 
any $j,l$ and $k=i\not=j$ we extract $q_{jl}$. Similar remarks hold for 
$P^{jl}$. In fact there is little advantage of sticking to the 
special smearing fields 
$f_{ab}\;^{cd}=f_{[a}^{(c}\; \delta_{b]}^{d)}$ and 
$g^{ab}\;_{cd}=g^{[a}_{(c}\; \delta^{b]}_{d)}$ and thus in what follows 
we will consider general 
$f_{ab}\;^{cd},\;\;g^{ab}\;_{cd}$ which are antisymmetric in $a,b$ and
symmetric in $c,d$.

\subsection{First steps: Geometrical operators}
\label{s5.3}

We proceed by supplying some details of the corresponding representations 
and the action of their corresponding Weyl algebras.
Consider e.g. the $(E,K)$ polarisation first. The vacuum 
satisfies $<E,g>_S\Omega=0$ and excited states are polynomials of 
Weyl elements $w_f(C)=\exp(-i\;<K,f>_C)$ applied to $\Omega$. Here 
\be \label{5.11}
<E,g>_S=\sum_{j\ge a}\int_S\; dx_a\; E^a_j\; g^j,\;
<K,f>_C=\sum_{j\ge a}\int_C\; dx^a\; K_a^j\; f_j
\ee
Thus a monomial of Weyl elements depends on a form factor 
\be \label{5.12}
F^a_j(x)=\sum_{I=1}^N\; \int_{C_I}\; dy^a\; f^I_j(y)\; \delta(x,y),\; j\ge a
\ee
with $N$ finite and we write $w[F]=\exp(-i<K,F>)$ in that case. One has 
$<\Omega,w[F]\Omega>=\delta_{F,0}$ displaying the discontinuous 
character of the configuration Weyl elements and the non-separability 
of the Hilbert space which are generic for Narnhofer-Thirring 
representations while the momentum Weyl elements 
$w[G]=\exp(-i<G,E>)$ based on the form 
factors    
\be \label{5.13}
G_a^j(x)=\sum_{I=1}^N\; \int_{S_I}\; dy_a\; g_I^j(y)\; \delta(x,y),\; j\ge a
\ee
are continuous. The Weyl algebra is governed by the Poisson brackets
\be \label{5.14}
\{<E,g>(S),<K,g>_c\}={\rm G_N}\;\sum_{p\in S\cap c},\; 
\sum_{j\ge a}\; (f^j\; g_j)(p)\;\sigma_p(S,c)\;\kappa_p^a(S,c),\;\;\;
\kappa_p^a(S,c):=[\frac{n^S_a \dot{c}^a}{|\sum_b n^S_b \dot{c}^b|}]_p
\ee
where it is understood that the last fraction is only computed 
for $\sigma_p(S,c)\not=0$ otherwise the contribution from $p$ is dropped
and ${\rm G_N}$ is Newton's constant which in $D+1$ spacetime 
dimensions relates to the Planck length as 
$\hbar {\rm G_N}=:\ell_P^{D-1}$. 

It follows that the the states $w[G]\Omega$ are eigenstates of the 
$E[F]$ with eigenvalue $\ell_P^2 <F,G>$ 
where $\ell_P^2$ is the Planck area 
and thus geometrical operators such as 
areas and volumes have explicitly computable eigenvalues without 
having to perform complicated harmonic analysis on SU(2) as 
is necessary in LQG. We exemplify this for the area operator 
corresponding to a co-dimension 1 surface 
\be \label{5.14a}
{\rm Ar}[S]:=\int_{[0,1]^{D-1}}\; d^{D-1}u\; \sqrt{\det(S^\ast q)}
=\int_{[0,1]^{D-1}}\; d^{D-1}u\; 
\sqrt{\sum_j [\sum_{a\le j}\; n_a^S\; E^a_j]^2}
\ee
with $n_a^S(u)=\epsilon_{ab_1..b_{D-1}}\; 
S^{b_1}_{,u_1}..
S^{b_{D-1}}_{,u_{D-1}1}$. It
is quantised by exactly the same steps as in \cite{10}, generalised to 
any dimension. We therefore can be brief here.  
We note that the paths entering $w[G]\Omega$ can be decomposed into 
edges of a graph $\gamma$
which nowhere intersect except, possibly, at its endpoints.
One difference with LQG is that, because the Gauss constraint has been 
solved, the 
graph need not be closed, i.e. edges may end in 1-valent vertices.
Then one adapts the edges of that graph further into edges of type
inside, outside and up, down and transversal, co-planar respectively    
with respect to $S$ and notices that only edges contribute that are 
transversal. One further decomposes $S$ into pieces $\Box$ and notices that 
only those arbitrarily small area elements $\Box$ contribute which contain 
an intersection point $p$ with at least one transversal edge. Let 
$P(S,\gamma)$ be the set of intersections points of the edges of $\gamma$
intersecting $S$ transversally,  
let $T_p(\gamma)$ be the set of edges $C$ of $\gamma$ intersecting $S$ at 
$p$ transversally and let $f^C_j$ be the smearing functions corresponding to
$C$. Then the area eigenvalue is given by 
\be \label{5.15}
\ell_P^{D-1}\sum_{p\in P(S,\gamma)}\; 
\sqrt{\sum_j\;[\sum_{a\le j}\; \sum_{C\in T_p(\gamma)}\; 
\kappa^a_p(S,C)\; f^j_C(p)]^2}
\ee
Clearly the edge ``charges'' $f^j_C(p)$ which are in fact boundary values 
of functions $f^j_C$ along $C$ replace the spin quantum numbers of 
LQG. Note also that the function $\kappa^a_p(S,C)$ can take either sign and 
replaces the distinction between edges of up and down type of LQG. In contrast
to LQG, co-planar edges do not contribute to (\ref{5.15}). Note that 
in LQG area operators for intersecting surfaces do not commute while 
the (\ref{5.14a}) do.

We now supply the analogous information for e.g. the $(Q,p)$ polarisation
which is rather different from the LQG situation. First of all, a 
``graph'' is now no longer one dimensional but rather 2-dimensional 
and should rather be called a {\it membrane} $\gamma$. It will be convenient 
to decompose a membrane $\gamma$ into 2d 
faces $C$ which intersect at most in 
1-dimensional boundary lines $l$. Let
\be \label{5.16}
<Q,g>_S=\int_S\; dx_{ab}\; g^{ab}\;_{cd}\; Q^{cd},\;
<p,f>_C=\int_C\; dx^{ab}\; f_{ab}\;^{cd}\; p_{cd}
\ee
We take the convention that 
coordinates have dimensions of length, thus $Q$ is dimensionless 
while $K$ has dimension of inverse length. 
Thus a monomial of Weyl elements $\exp(-i<p,f>_C)$
depends on a form factor 
\be \label{5.17}
F^{cd}(x)=L^{-1}\;
\sum_{I=1}^N\; \int_{C_I}\; dy^{ab}\; (f^I)_{ab}\;^{cd}(y)\; 
\ee
with $N$ finite and we write $w[F]=\exp(-i<K,F>)$ in that case where 
$L$ is an arbitrary unit of length in order to keep 
$f^I$ dimensionfree. One has 
$<\Omega,w[F]\Omega>=\delta_{F,0}$ displaying again the discontinuous 
character of the configuration Weyl elements and the non-saparability 
of the Hilbert space which are generic for Narnhofer-Thirring 
representations while the momentum Weyl elements 
$w[G]=\exp(-i<G,Q>)$ based on the form 
factors    
\be \label{5.18}
G_{cd}(x)=\sum_{I=1}^N\; \int_{S_I}\; dy_{ab}\; 
(g_I)^{ab}\;_{cd}(y) (y)\; \delta(x,y)
\ee
are continuous. The Weyl algebra is governed by the Poisson brackets
\ba \label{5.19}
\{<Q,g>_S,<p,f>_C\} &=& {\rm G_N}\;\sum_{p\in S\cap C},\;
[\frac{n^{ab}_C\; N_{cd}^S}{|n^{gh}_C\; N_{gh}^S|}](p)\; 
[(f^C)_{ab}^{ef}\; (g^S)^{cd}_{ef}](p)
\nonumber\\
n^{ab}_C(t)&=&C^{[a}_{,t_1}(t) \;C^{b]}_{,t_2}(t),\;\;
C:[0,1]^2\mapsto \sigma
\nonumber\\
N_{ab}^S(u)&=& \epsilon_{ab c_1..c_{D-2}}\; 
S^{c_1}_{,u_1}(u).. \;S^{c_{D-2}}_{,u_{D-2}}(u),\;\;
S:[0,1]^{D-2}\mapsto \sigma
\ea
where the sum is over intersection points $p$ of $S$ and $C$ which are 
transversal, i.e. $T_p(C)\oplus T_p(S)=T_p(\sigma)$ i.e. the tangent 
spaces of $C,S$ at $p$ span the whole tangent space which makes sure 
that the denominator in (\ref{5.19}) is non-vanishing. Note that all 
indices take full range and the summation convention applies.

To compute the ``area'' eigenvalue requires a bit more work because 
there is no direct analogue in LQG both because we work in general $D$ and 
because we deal with different smearing dimensions of the configuration 
degrees of freedom. The natural area in this polariation is the 
induced volume of $D-2$ surfaces $S$ which has co-dimension 2 rather than 
1. We will refer to the (D-2)-manifold $S$ as co-surface and to its 
induced volume as area. The area functional is given by 
\be \label{5.20}
{\rm Ar}[S]:=\int_{[0,1]^{D-2}}\; d^{D-2}u\; \sqrt{\det(S^\ast q)}
=\int_{[0,2]^{D-1}}\; d^{D-2}u\; 
\sqrt{N^S_{ab}\; N^S_{cd}\; Q^{ac}\; Q^{bd}}
\ee
To quantise (\ref{5.20}) we partition $S$ into cells $\Box$ and write 
it as a Riemann sum over cell contributions. In order to write it in terms 
of the elementary variables $<Q,g>_\Box$ we have to identify suitable 
tensors $g$. Consider the diffeomorphism invariant tensors 
\be \label{5.21}
[g_r\;^s]^{ab}\;_{cd}:=\delta^{[a}_r\; \delta^{b]}_{(c}\; \delta^s_{d)}
\ee
then it is not difficult to see that (\ref{5.20}) becomes 
\be \label{5.21}
{\rm Ar}[S]=\lim_{{\cal P}\to S}\; \sum_{\Box \in {\cal P}}\;
\sqrt{-\; \sum_{r,s}\; <Q,f_r\;^s>_\Box\;\;
<Q,f_s\;^r>_\Box}
\ee
where $\cal P$ is a partition of $S$ and the limit indicates that the 
partition approaches the continuum. 

Interestingly, the argument 
of the square root in (\ref{5.20}), (\ref{5.21}) is not 
positive for general symmetric $Q^{ab}$, it is only when $Q^{ab}$ is 
positive which is classically the case by definition. 
In the quantum theory, as we will not impose positivity 
on general states but just on semiclassical ones, we will thus introduce a 
modulus into (\ref{5.21}) in the sense of the spectral theorem (in LQG
one proceeds similarly with the volume operator defined by the densitised 
triad which involves a square root of a quantity which is only positive 
for triads of positive orientation).
The rest of the derivation is then very similar to the situation 
before. Eventually only those $\Box$ contribute which contain 
transversal intersection points $p$ of $S$ and the faces $C$ of the 
membrane $\gamma$. Let $P(S,\gamma)$ be the set of those and for 
$p\in P(S,\gamma)$ let
$T_p(\gamma)$ be the set of faces $C$ intersecting $S$ transversally in $p$
in an adapted decomposition of $\gamma$ into tranversal faces and the 
rest. Then the area eigenvalue is given by 
\ba \label{5.22}
&&\frac{\ell_P^{D-1}}{L}\; \sum_{p\in P(S,\gamma)}\;
\sqrt{|- N^S_{ab}\; F_S^{bc}\; N^S_{cd}\; F_S^{da}|(p)}
\nonumber\\
F_S^{cd}(p) &=&
\sum_{C\in T_p(\gamma)}\; 
\frac{n^{ab}_C\; (f^C)_{ab}\;^{cd}}{|n^{ef}_C\; N^S_{ef}|}(p)
\ea
where we used as above the spectral theorem i.e. that the
operators $<Q,g>_\Box$ are diagonal on the $w[G]\Omega$. Again the 
analogy with LQG is quite obvious, the spins have been replaced 
by the smearing tensors $f^C$ and intersection numbers by $n_C, N^S$
tensors. The co-surface area operators are mutually commuting even for 
intersection co-surfaces.

\subsection{Quantum dynamics and inverse volume operator}
\label{s5.4}

We will explore these new representations in more detail in 
forthcoming papers. If we do not work in the $(E,A)$ polaristaion 
then the Ricci scalar of $q$ contributing to the Hamiltonian
constraint is not accessible via the curvature
$F$ of $A$ \cite{11} and has to be defined directly. A direct definition 
has also been given in \cite{25}. However, with the Gauss constrained 
removed this is much easier and the corresponding curvature operator 
will commute with all other operators that depend just on the D metric.
In particular, one can define inverse powers of the metric determinant 
using the analogs to \cite{11} of Poisson brackets between the volume 
functional and $K$ or $p$. In that sense much of the technology developed for 
the $(A,E)$ polarisation with quantum Gauss constraint solution can be 
transferred to the $(K,E),\; (e,P),\; (p,Q),\; (q,P)$ polarisation with 
classical Gauss constraint solution. 

An alternative to defining inverse powers of the metric determinant which 
is even practically feasible arises as follows: Note that in the UTPD
and using the corresponding D-bein $e$ the metric determinant is just given by 
$\det(q)=\prod_{a=1}^D\; [e_a^{j=a}]^2$. The density weight one 
Hamiltonian 
constraint is a polynomial (e.g. of degree at most ten in $q_{ab}, P^{ab}$
for $D=3$ and in vacuum) 
up to a factor $\sqrt{\det(q)}^{-n}$ ($n=5$ for vacuum and $D=3$)
which also renders the Ricci scalar
contribution polynomial. It follows that the notoriously 
arising inverse powers of $\sqrt{\det(q)}$ can be tamed if we can tame 
inverse powers of the $X_a(x):=|e_a^{j=a}(x)|$ as operators. Such a taming 
can be systematically addressed using constructive renormalisation, i.e. 
introducing IR (compact $\sigma$) and UV (lattice) cut-offs so that 
one is dealing with finite dimensional quantum mechanical systems when 
both regulators are installed for which we can choose e.g. a Schroedinger 
representation. We then are confronted with the question 
how to give meaning to inverse powers of the $X_a(m)$ where $m$ denotes a 
lattice point of the lattice (say with $N^D$ vertices and the off diagonal 
components of $e_a^j$ appear only polynomially; we understand that the 
$e_a^j$ and $P^a_j$ are properly smeared in $1,D-1$ dimensions with respect 
to the edges and faces of the lattice and dual cell comple respectively). 
Surprisingly, there exists 
an explicit dense domain ${\cal D}$ in $L_2(\mathbb{R}^{n(D,N)},d^{n(d,N)}e)$ 
where 
$n(D,N)=\frac{D(D+1)}{2}\; N^D$ is the number of multiplication operators 
$e_a^j(m),\; a\le j$ which is 1. invariant under any polynomial in the 
$\partial/\partial e_a^j(m)$ and {\it any integer power} of any $e_a^j(m)$
and 2. such that the norms of the vectors 
in the image of such operator on $\cal D$ can be analytically 
evaluated in closed form
\cite{33}. The proof of existence of such a domain was sketched 
already in \cite{34} (albeit using a singular value rather than an 
upper triangular decomposition and without solving the Gauss constraint)
and claimed in \cite{35}. The complete details can be found in \cite{33}. 
This domain is also helpful in the $(q,P)$ polarisation but evaluation 
e.g. of $||\sqrt{\det(q)}^{-n}\psi||,\; \psi\in {\cal D}$ which is granted 
to be finite, appears to be analytically more involved. 
As stressed in \cite{35} the domain $\cal D$ will be of immediate relevance 
without renormalisation for applications in quantum cosmology including 
backreaction where inverse powers of the quantum scale factor $a$ can now 
be tamed and computations be done in closed form, the classical big bang 
singularity $a=0$ poses no problem on $\cal D$ and one can work with 
a Schr\"odinger representation rather than using the representation used 
in Loop Quantum Cosmology (LQC) \cite{36} without encountering the 
quantisation ambiguities of LQC while the Hamiltonian constraint 
operator remains well defind when propagating states through the classical 
singularity \cite{37}.

Yet another exciting possibility connected with these representations is 
that Weyl states based on smearing functions along infinite graphs 
are part of the Hilbert space and do not require the infinite tensor product
extension \cite{26}. Moreover, they are not required to be closed. For 
instance, for non-compact $\sigma=\mathbb{R}^D$ or compact $T^D$ 
we can consider the countable, dense graph based on the coordinate lines 
into $a=1,..,D$ direction at constant 
$x^b=N^b+\frac{k^b}{2^{n^b}},\; b\not=a,
N^b\in \mathbb{Z},\; n^b\in \mathbb{N},k^b=1,3,..,2^{n^b}-1$ which fill 
$\sigma$ densely and can let $K_a^j,\;j\ge b$ be smeared along those 
lines e.g. with smooth functions of rapid decrease. These Weyl functions   
represent non-degenerate quantum geometries as they are $V(O)$ eigenstates 
with non-vanishing eigenvalue where $V$ is the volume operator 
of an open region $O\subset \sigma$. It is therefore much simpler 
to construct non-degenerate states than in LQG where non-degeneracy can 
only be obtained by taking contable superpositions of states 
supported on finite graphs which we explore in the appendix. 
It is on such states that we expect 
the hypersurface deformation algebra of GR to be represented rather 
than on dgenerate states because that algebra requires non-degeneracy 
as part of its definition \cite{13}.

\section{Conclusion and Outlook}
\label{s6}

In this paper we have shown that the Gauss constraint of the internal 
gauge group of GR that is necessary when fermionic matter is present,
can be solved classically using a perfect gauge that we have coined 
upper triangular positively diagonal (UTPD). Like the unitary gauge 
of the electroweak interaction it can be used to get rid of the spurious 
degrees of freedom that are encoded in a Vielbein. 

With that gauge group being absent, the motivation for a polarisation 
of the gravitational phase space in terms of connections is largely absent
thereby reviving interest in two more natural polarisations, one 
based on UTPD D-beins and the other based on the ADM D-metrics and their 
corresponding relatives. These polarisations are naturally available 
in any spatial dimension D and do not require higher gauge groups and 
simplicity constraints. The UTPD Vielbein formulation uses D-beins
smeared in 1 dimensions and their conjugate momenta smeared in D-1 dimensions 
while the metric formulation uses D-metrics smeared in 2 dimensions and 
their conjugate momenta in D-2 dimenions. Corresponding rigorous 
representations of the Weyl algebra can be chosens for instance of 
of Narnhofer-Thirring type \cite{22} and are such that the D-bein or
the D-metric annhihilates the vacuum. 

The UTPD formulation is most suitable in situation wheen all constraints 
have been gauge fixed as the UTPD Weyl algebra does not display a simple 
behaviour e.g. under spatial diffeomorphisms. The D-metric formulation 
is also suitable in situations in which the Hamiltonian and 
spatial constraints have not been gauge fixed and in which case one 
works with manifestly gauge invariant and diffeomorphism covariant
quantities. In either case the hypersurface deformation algebra is 
unchanged on the reduced phase space.

Among the advantages of working with the Gauss gauge fixed formalism
besides being readily available in any D is the fact that i. no 
non-commutative structure appears even at the classical level 
\cite{9}, ii. there is much more flexibility in the way we choose 
to smear the fields with test functions and thus to pick the Weyl
algebra, iii. non-degerate states, whose importance for a proper 
implementation of the hypersurface deformation algebra has recently been 
stressed \cite{13}, are much simpler to construct, because the 
volume operator is diagonal in all representations sketched above  
and iv. for the same reason, the quantum dynamics is much simpler 
as geometric quantities built from the D-metric are mutually commuting,
the volume operator is under complete control 
and otherwise all the techniques invented for the D=3 polarisation $(A,E)$ 
\cite{11} apply verbatim.

\begin{appendix}

\section{Non-degenerate LQG states}
\label{sa}

In the first subsection we motivate a possible definition of 
non degenerate states. In the second we show their existence in the 
LQG Hilbert space. In the third we establish some of their properties,
in particular that while they do not form a linear subspace, the set 
of non-degenerate states contains a cone the span of whose elements 
is dense. In the fourth we introduce non-degenerate distributions 
which may be used to construct physically interesting 
{\it non-degenerate habitats} and in the fifth we 
perform a preliminary investigation
using coherent states to analyse whether a representation of $\mathfrak{h}$ 
on such a habitat or even within the LQG 
Hilbert space is conceivable in the non-degenerate sector. 
Our investigations will have a 
pioneering character, are far from complete at the moment and provide 
ample opportunities for refinement and generalisation. At the very least,
they provide additional motivation to study LQG renormalisation \cite{18}.

\subsection{Definition}
\label{sa.1} 

Intuitively, a non-degenrate quantum state is such that the metric 
operator $q_{ab}$ is ``everywhere invertible'' on $\sigma$. A classically
equivalent condition is that $\det(q)$ is ``everywhere'' positive or 
still equivalent that the volume $V(R)$ is positive for ``every'' 
region $R$ in $\sigma$. In this form, the condition can be meaningfully 
transferred into the quantum theory of LQG because $V(R)$ is a well 
defined operator. 

A classical state is just a point in the classical phase space.
The closest quantum analog of a classical state is a quantum state.
If we do not want to leave the folium of the LQG representation then 
a quantum state is a density matrix $\rho$ in the LQG Hilbert space. 
The simplest
density matrices are of rank one $\rho=\psi \; <\psi,.>_{{\cal H}},\; 
||\psi||=1$, hence we start with those. These are pure states in the 
LQG Hilbert space as the LQG representation is irreducible. Our first try
is therefore to say that a pure state $\psi\in {\cal H}$ is non-degenerate 
if $<\psi,\;V(B)\psi>\;>0$ for every {\it measurable} (Borel) subset 
$B\subset \sigma$. However, this condition is too strong because 
$\psi$ is in the Cauchy completion of the SNWF thus depends on an most
countable set $P$ of vertices. Since the countable point sets of a Borel 
$\sigma$ algebra are measurable 
and have Lebesgue measure zero, 
with $B$ also $B-P$ is measurable and thus 
$<\psi,\;V(B-P)\psi>=0$. The following is a working definition.
\begin{Definition} \label{defa.1} ~\\
A pure state $\psi\in {\cal H},\; ||\psi||=1$ is called non-degenerate iff
it is in the domain of $V(O)$ and 
\be \label{a.1}      
<\psi,\; V(O)\;\psi>\;>0
\ee
for every open subset $O\subset \sigma$. The set of non-degenerate 
pure states is denoted by ${\cal M}_{{\rm ND}}$.
\end{Definition}

\subsection{Existence}
\label{sa.2}

Let $P$ be the countable set of vertices on which $\psi$ depends. Then
$P\cap O\not=\emptyset$ for any open set $O$ in order that (\ref{a.1}) 
holds. It follows that every open set $O$ must contain an {\it infinite 
number} 
of points of $P$. Thus it is not at all clear that non degenerate 
(normalisable) states exist at all in the LQG Hilbert space or 
that any such state should be in the domain of the volume 
operator for any region. The following 
consideration shows that the set of non-degenerate states in the 
LQG Hilbert space is at least countably infinite. For simplicity, we consider 
$\sigma=\mathbb{R}^3$ in what follows, for general $\sigma$ one uses 
the usual patching of coordinates.
\begin{Example} \label{exa.1} ~\\ 
Let $\gamma_0$ be a closed graph with one 4-valent, non-coplanar 
vertex at the origin $o$, and another 4-valent co-planar
vertex. 
Consider the closed subspace of $\cal H$ obtained from the span 
of SNWF over $\gamma_0$ and diagonalise $V(R), o\in R$ in that subspace.
Label the corresponding countably infinite 
number of eigenstates with non-vanishing eigenvalue 
$\lambda_I>0$ by $T_0^I,\; I\in \mathbb{N}$. Let 
$\nu:\; \mathbb{N}_0\to \mathbb{Q}^3$ (rational numbers considered as a 
subset of $\sigma=\mathbb{R}^3$) be an arbitrary 
bijection with $\nu(0)=o$ 
and $T^I_n$ the parallel translate of $T^I_0$ from $o=\nu(0)$ 
to $\nu(n)$.
Let 
\be \label{a.2}
\psi^I:=\kappa\; \sum_{n\in \mathbb{N}_0}\; [1+n]^{-1}\; T^I_n,\;\;
\kappa^2:=[\sum_{n=0}^\infty\;[1+n]^{-2}]^{-1}
\ee
Then every $\psi^I$ is a pure non degenrate state. Moreover
\be \label{a.3}
<\psi^I,\; \psi^J>=\delta^{IJ},\;\;
0<\; <\psi^I,\; V(O)\; \psi^I>\;\le \lambda_I,\;\;
<\psi^I,\; V(O)^2\; \psi^I>\;\le \lambda_I^2<\infty
%,\;
%c:=\frac{\sum_{n=1}^\infty n^{-2}}{\sum_{n=1}^\infty n^{-4}
\ee    
In particular every $\psi^I$ is in the domain of every $V(O)$.
\end{Example}
\begin{proof}:\\
Since SNWF are orthogonal as soon as their graphs differ in at least one 
point, it is clear that $<T^I_m,\; T^J_n>=\delta^{IJ}\;\delta_{mn}$. 
It follows 
\be \label{a.4}
<\psi^I,\psi^J>=\kappa^2\;\delta^{IJ}\; \sum_{n=0}^\infty \; [1+n]^{-2}
=\delta^{IJ}
\ee
Next, since the volume operator supported in $O$ 
acts on a SNWF vertex wise if that vertex is in $O$ and otherwise 
annihilates it 
\be \label{a.5}
V(O)\psi^I
=\kappa\; \sum_n\; [1+n]^{-1}\; V(O)\;T^I_n
=\lambda_I\; \kappa; \sum_{\nu(n)\in O}\; [1+n]^{-1}\; \;T^I_n
\ee
whence
\be \label{a.6}
0<\; <\psi^I\;V(O)\psi^I>=\lambda_I\; 
\frac{\sum_{\nu(n)\in O} [1+n]^{-2}}{\sum_{n\in \mathbb{N}_0}\; [1+n]^{-2}}
\le \lambda_I
\ee
and similar for $V(O)$ replaced by $V(O)^2$ shows that $\psi^I$ is in the 
domain of every $V(O)$.
\end{proof}
Clearly one can show the the set of non degenerate states even has 
countably infinite cardinality by manipulating the graph label rather than 
the spin and intertwiner labels of the SNWF. The mechanism at work 
in (\ref{a.6}) is the following: As $\mathbb{Q}^3$ is dense in 
$\mathbb{R}^3$, the sum in the numerator is over the set of points 
$p\in O\cap \mathbb{Q}^3$ which is countably infinite. However, each 
of these points is given some finite label $n\in \mathbb{N}$ and the sum 
is weighted by $n^{-2}$ which makes it converge even when summing over all
of $\mathbb{N}$.     

As just mentioned, 
the vertex set $P$ of $\psi^I$ here coincides with $\mathbb{Q}^3$ for each $I$.
Any other set $P$ which is dense in $\mathbb{R}^3$ would serve the same 
purpose. This gives us the possibility to distribute the points in $P$
in a better controlled way. For instance consider the map
\be \label{a.7}
\nu(N,n,k):=N+\frac{k}{2^n},\; N\in \mathbb{Z},\; n\in \mathbb{N},\;
k\in M_n=\{1,3,5,.., 2^n-1\}
\ee
This is easily shown to be an injection into $\mathbb{R}$ which is dense 
because the points at fixed $n$ partition $\mathbb{R}$ into intervals of 
length $2^{-(n-1)}$. We can extend (\ref{4.7}) to $\mathbb{R}^3$ by 
\be \label{a.8}
\nu(N,n,k):=
N+\frac{k}{2^n},\; N\in \mathbb{Z}^3,\; n\in \mathbb{N},\;
k\in M_n^3
\ee
Consider now any finite graph $\gamma_0$ with a vertex in the origin 
$o$ and a state $\psi_0,\; ||\psi_0||=1$ 
in the closed linear span of SNWF over $\gamma_0$
with the property that for each vertex $v$ in the vertex set $V$ 
of $\gamma_0$ and any open set $O_v$ with $V\cap O_v=\{v\}$ we have 
$<\psi_0,\; V(O_v)\; \psi_0>>0$. Thus $\psi_0$ has positive volume expectation  
value at every $v\in V$ but is not necessarily an eigenstate of 
any of the $V(O_v)$. Let $\psi_{N,n,k}$ be the parallel translate 
of $\psi_0$ from $o$ to $\nu(N,n,k)$ and $\Psi$ the state 
\be \label{a.9}
\Psi:=\kappa\; \sum_{N,n,k}\; [\prod_{a=1}^3\;[1+N_a^2]^{-1/2}]\;n^{-1}\;
2^{-3(n-1)/2} \; \psi_{N,n,k},\;\;
\kappa^{-2}=
\sum_{N,n,k}\; [\prod_{a=1}^3\;[1+N_a^2]^{-1}]\;n^{-2}\; 2^{-3(n-1)} 
\ee
Then $||\Psi||^2=1$ because the $\gamma_{N,n,k}$ are mutually different 
thanks to $\nu$ being an injection whence $<\psi_{N,n,k},\;
\psi_{N',n',k'}>=\delta_{NN'}\;\delta_{nn'}\;\delta_{kk'}$ and 
$|M_n|=2^{n-1}$ was used. Let us label the vertices of $\gamma_0$ by $v^1,..,v^M$
and set $V^l:=<\psi_0, V(O_{v^l})\psi_0>,\; l=1,..,M$. Let 
$v^l_{N,n,k}$ be the translate of $v^l$ and given $O$ 
consider the sets $M^l(O)=\{(N,n,k);\; v^l_{N,n,k}\in O\}$.
Then
\be \label{a.10}
0<\; <\Psi, V(O)\Psi>=\kappa^2\; \sum_{l=1}^M\; V^l\; 
\sum_{(N,n,k)\in M^l(O)}\;
\sum_{N,n,k}\; [\prod_{a=1}^3\;[1+N_a^2]^{-1}]\;n^{-2}\; 2^{-3(n-1)} 
\le \sum_{l=1}^M V^l
\ee
where it was used that $V(O)$ only changes the intertwiner structure 
of a SNWF but neither spins nor graph so that also 
$<\psi_{N,n,k},V(O)\psi_{N',n',k'}>$ is non-vanishing 
iff $N=N',\;n=n',\; k=k'$. If $V(O)$ is replaced by $V(O)^2$ in 
(\ref{a.10}) then $V^l$ gets replaced by $[V^l]^2$.

To see explicitly that $\Psi$ is in the 
Cauchy completion, we introduce the sequence of states 
\be \label{a.11}
\Psi^{N_0,n_0}:=
\kappa^2\; \sum_{|N_1|,|N_2|,|N_3|\le N_0}\; 
\sum_{n=1}^{n_0}\;\sum_{k\in M_n^3} [1+||N||^2]^{-3/2} \; n^{-1}\;
2^{-3(n-1)/2}\; \psi_{N,n,k}
\ee
Then say for $N_0> M_0, n_0> m_0$ (other possibilities similar)
\ba \label{a.12}
&& ||\Psi^{N_0,n_0}-\Psi^{M_0,m_0}||^2
=\kappa^2\{
\sum_{n=m_0+1}^{n_0}\; \sum_{|N_1|,|N_2|,|N_3|\le N_0} 
\nonumber\\
&& +\sum_{n=1}^{m_0}\;[  
3\sum_{M_0<|N_1|\le N_0} \; \sum_{|N_2|,|N_3|\le M_0}
+3\sum_{M_0<|N_1|,|N_2|\le N_0} \; \sum_{|N_3|\le M_0}
+\sum_{M_0<|N_1|,|N_2|,|N_3|\le N_0}]
\nonumber\\
&& n^{-2}\; [\prod_{a=1}^3\;[1+N_a^2]^{-1}]\}
\nonumber\\
&\le &
\kappa^2\;\{
[\sum_{n=m_0+1}^{n_0}\; n^{-2}]\;
[\sum_{N\in \mathbb{Z}} [1+N^2]^{-1}]^3\; 
\nonumber\\
&& +3[\sum_{n\in \mathbb{N}}\; n^{-2}]\;
[\sum_{N\in \mathbb{Z}} [1+N^2]^{-1}]^2
[\sum_{M_0<N\le N_0} [1+N^2]^{-1}]
\nonumber\\
&&+3[\sum_{n\in \mathbb{N}}\; n^{-2}]\;
[\sum_{N\in \mathbb{Z}} [1+N^2]^{-1}]
[\sum_{M_0<N\le N_0} [1+N^2]^{-1}]^2
\nonumber\\
&&+[\sum_{n\in \mathbb{N}}\; n^{-2}]\;
[\sum_{M_0<N\le N_0} [1+N^2]^{-1}]^3
\}
\nonumber\\
&\le &
\kappa^2\;\{
[\sum_{n=m_0+1}^{n_0}\; n^{-2}]\;
[\sum_{N\in \mathbb{Z}} [1+N^2]^{-1}]^3\; 
\nonumber\\
&& +7[\sum_{n\in \mathbb{N}}\; n^{-2}]\;
[\sum_{N\in \mathbb{Z}} [1+N^2]^{-1}]^2
[\sum_{M_0<N\le N_0} [1+N^2]^{-1}]
\}
\ea
and 
\be \label{a.13}
\sum_{m_0<n<n_0} \; n^{-2}
\le \sum_{n=m_0}^\infty \; n^{-2}
\le m_0^{-2}[\sum_{r\in \mathbb{N}}\;\sum_{l=0}^{m_0-1} \; 
[\frac{m_0}{r m_0+l}]^2
\le m_0^{-2} m_0\; \sum_{r=1}^\infty r^{-2}=\frac{c}{m_0}
\ee
and 
\be \label{a.14}
[\sum_{M_0<N\le N_0} [1+N^2]^{-1}]
\le 2 \sum_{N=M_0}^\infty N^{-2} 
\le 2\; c \; M_0^{-1}
\ee
Thus $l\mapsto \Psi^{N_0(l),n_0(l)}$ is a Cauchy sequence for any choice 
of diverging sqequences $N_0(l),n_0(l)$. Here $N_0,n_0$ respectively have 
the physical meaning of IR and UV cut-off respectively.

\subsection{Properties}
\label{sa.3}

To systematise the construction of non-degenerate states we recall that 
the volume operators $V(O)$ mutually commute for any choice of $O$ and
have pure point spectrum. They can therefore be simultaneously 
diagonalised with pure point spectrum. Accordingly, by the spectral theorem 
(which also holds in non separable Hilbert spaces) there exists a function
$\Lambda:\; {\cal O}\to \mathbb{R}^+_0$ from the open sets in $\sigma$ to 
the non-negative reals and normalisable elements $E_\Lambda\in {\cal H}$ 
such that $V(O)\; E_\Lambda=\Lambda(O)\; E_\Lambda$. Non-degenerate 
eigenstates are specified by those functions $\Lambda$ with 
$\Lambda(O)>0$ for every $O$. 

We will determine the complete set of all  
$E_\Lambda$ and show that 
no non degenerate eigenstates exist. We benefit from 
the large amount of information that is available about the action of 
$V(O)$ on SNWF $T_s,\; s=(\gamma,j,\iota)$. 

We know that 
\be \label{a.14a}
V(O)\;T_s=\sum_{v\in O\cap V(\gamma)}\; V_v\; T_s
\ee
where $V_v$ is a positive s.a. operator that preserves $\gamma,j$ 
and just changes 
$\iota_v$, the intertwiner. Since for fixed $\gamma,j$ the set of linearly 
independen $\iota_v$ intertwiners is finite dimensional just depending 
on the intersection structure and the spin labels of the edges adjacent to 
$v$, we may diagonalise $V_v$ and obtain a different ONB which one 
call the {\it volume network functions} (VNWF). We label them as 
$T_i,\; i=(\gamma,j,\lambda)$ where $\lambda$ is the collection of 
eigenvalues $\lambda^v$ including multiplicity. Then, while diagonalisation 
of $V_v$ is a computational challenge in finite dimensional linear 
algebra, it is in principle clear how to find the $T_i$. It follows
\be \label{a.15}
V(O)\; T_i=\Lambda(O,i)\; T_i,\;\Lambda(O,i)=\sum_{v\in V(\gamma)\cap O}
\lambda^v,\;i=(\gamma,j,\lambda)
\ee
Thus the $T_i$ are simultaneous eigenstates of all $V(O)$ but certainly 
they are all degenerate because for every fixed $i$ and thus fixed 
$\gamma_i$ we find infinitely many $O$ such that $O\cap V(\gamma_i)=
\emptyset$. 

We proceed to compute {\it all} simultaneous eigenstates 
$E_\Lambda$ of all $V(O)$. 
\begin{Definition} \label{defa.2} ~\\
i.  The uncountably infinite 
index set of all VNW is denoted by $\cal I$.\\
ii. For each open subset $O$ the uncountable subset $I(O)\subset {\cal I}$ 
consists of those $i$ with $\Lambda(O,i)>0$.\\
iii. Let $T_z\in {\cal H}$ be a linear combination  
\be \label{a.16}
T_z:=\sum_{i\in {\cal I}}\; z_i\; T_i,\; 
||T_z||^2=\sum_{i \in {\cal I}}\; |z_i|^2<\infty
\ee
The support $I$ of $T_z$ is the countable subset $I\subset {\cal I}$ 
consisting of those $i$ with $z_i\not=0$.\\
iv. $T_z$ is non-degenrate iff for all $O\in {\cal O}$  
we have 
a. $I\cap I(O)\not=\emptyset$ 
and b. $T_z$ is in the domain $V(O)$, that is 
\be \label{a.16a}
0<\sum_{i\in I\cap I(O)}\; |z_i|^2\; V(O,i)^2<\infty
\ee
\end{Definition}
\begin{Theorem} \label{tha.1} ~\\
Let $E_\Lambda$ be a common eigenvector for all $V(O)$ with respective 
eigenvalue $\Lambda(O)$. Then $E_\Lambda$ is of the form $T_z$ and  
there exists a finite set of points $V$ (possibly empty)
such that for $i$ in the support $I$ of $T_z$ we have $V\subset V(\gamma_i)$
(vertex set).
If $v\in V$ then $\lambda^v_i=\lambda^v$ is independent of $i$ 
and can be positive. If $v\in V(\gamma_i)-V$ then $\lambda^v_i=0$.
The eigenvalue is given by 
\be \label{a.17}
\Lambda(O)=\sum_{v\in O\cap V}\; \lambda_v
\ee
\end{Theorem}  
\begin{proof}:\\
As VNWF form an ONB of $\cal H$ we 
can write a general eigenstate $E_\Lambda$ always 
in the form (\ref{a.16}).
Then the eigenvalue equation becomes 
\be \label{a.18}
\Lambda(O)\; E_\Lambda=\sum_{i\in I}\; z_i\; \Lambda(O,i)\; T_i
\ee
which implies 
\be \label{a.19}
z_i \; [\Lambda(O)-\Lambda(O,i)]=0
\ee
for all $i$ which for $i\in I$ enforces $\Lambda(O)=\Lambda(O,i)$, that
is, $\Lambda(O,i)$ does not depend on $i\in I$ for any $O$. Consider 
$i,i'\in I,\;i\not=i'$. Using (\ref{a.15}) and by choosing $O$ to contain 
only one element 
of $V(\gamma_i)\cup V(\gamma_{i'})$ we find that
necessarily $\lambda^i_v=0$ for $v\in V(\gamma_i)-V(\gamma_{i'})$ and 
$\lambda^{i'}_v=0$ for $v\in V(\gamma_{i'})-V(\gamma_i)$ and 
$\lambda^i_v=\lambda^{i'}_v$ for 
$v\in V(\gamma_i)\cap V(\gamma_{i'})$. 

It follows that there exists 
a common subset $V$ (possibly empty) of all $V(\gamma_i),\; i\in I$ 
at which $\lambda^i_v=\lambda_v$ 
is the same for each $i\in I$ and can be positive while 
$\lambda^i_v=0$ for all $v\in V(\gamma_i)-V$. Formula (\ref{a.17}) then 
follows from (\ref{a.15}). \\
\end{proof}
Examples of such states are zero 
volume states ($V$ is empty or all $\lambda^v=0,\;v\in V$) 
or single element states ($I$ consists
only of one element) or several element states ($I$ has more 
than one element) such that the $\lambda^i_v$ for $v\in V$ are all
equal e.g. because the eigenvalue $\lambda^v$ has multiplicity 
larger than unity for the same $j=j_i=j_{i'}$ or because different 
$j_i\not=j_{i'}$ still allow for the same intertwiner space using 
recoupling theory of angular momentum or because the graphs 
$\gamma_i\not=\gamma_{i'}$ are different but have the same $V$ and 
equal eigenvalues there (e.g. diffeomorphic states with 
fixed points in the vertices). The general case is a mixture of those.

Thus, there is a rich set of
volume eigenstates vor all $V(O)$ more general than the $T_i$. 
However, none of them is non-degenerate as 
the common susbset $V$ consists only of finitely many points. There 
is therefore no advantage to consider these more general states and we 
will keep working with the $T_i$ which are under better analytical
control.

A non-degenerate state is still of the form (\ref{a.16}). 
\begin{Theorem} \label{tha.2} ~\\
i. The set of non-degenrate states ${\cal M}_{{\rm ND}}$ is not a linear 
subspace of  $\cal H$.\\
ii. ${\cal M}_{{\rm ND}}$ contains a cone ${\cal C}_{{\rm ND}}$.\\
iii. The span 
${\cal D}_{{\rm ND}}$ of ${\cal C}_{{\rm ND}}$ is dense in $\cal H$.\\
iv. ${\cal M}_{{\rm ND}}-{\cal C}_{{\rm ND}}\not=\emptyset$.
\end{Theorem}
\begin{proof}:\\
i.\\ 
Suppose $\psi,\psi'$ are non-degenerate. Then we can write them as 
$\psi=T_z,\; \psi'=T_{z'}$ where $I\cap I(O),I'\cap I(O)\not=\emptyset$ 
for all $O$
and $I,I'$ are the supports of $\psi,\psi'$. Pick $I'=I$ and $i_0\in I$ 
and $z'_{i_0}=z_{i_0},\;z'_i=-z_i,\;i_0\not=i\in I$. Then 
$\psi+\psi'=T_{z+z'}=2 z_{i_0} T_{i_0}$ has support $I_0=\{i_0\}$ and thus 
$I_0\cap I(O)=\emptyset$ for any $O$ such that 
$O\cap V(\gamma_{i_0})=\emptyset$.\\
ii.\\
We define the cone ${\cal C}_{{\rm ND}}$ to consist of those non-degenrate
$T_z$ such that $z_i>0$ for $i\in I$ (i.e. only 
real, non-negative coefficients are allowed). Then for $s,s'>0$ we 
have $s\; T_z+s'\; T_{z'}=T_{s\;z+s'\;z'}$. Certainly 
$||T_{sz+s' z'}||\le s\;||T_z||+s'\;||T_{z'}||<\infty$ and  
$||V(O) T_{sz+s'z'}||\le 
s\;||V(O) T_z||+s'\;||V(O)T_{z'}||<\infty$ for all $O$ and 
\ba \label{a.20}
&& <T_{sz+s'z'},V(O)\;T_{sz+s'z'}> =
\sum_{i\in {\cal I}}\;|s z_i+s' z'_i|^2 \; \Lambda(O,i)
= s^2\; <T_z,V(O) T_z>+[s']^2\; <T_{z'},V(O) T_{z'}>
\nonumber\\
&& +2s\;s'\;\sum_{i\in I\cap I'\cap I(O)}\;z_i\;z'_i \; 
\Lambda(O,i)
\ge s^2\; <T_z,V(O) T_z>+[s']^2\; <T_{z'},V(O) T_{z'}> >0
\ea
thanks to the positivity of all involved numbers.\\
iii.\\
It suffices to show that for each $i\in {\cal I}$ and $\epsilon>0$ there
exists $M\in \mathbb{N}$, elements $T_{z^{(N)}},\; N=0,.,M$ in the 
cone ${\cal C}_{{\rm ND}}$ and complex coefficients $c_N,\; N=0,..,M$ 
such that $||T_i-\sum_{N=0}\; c_N\; T_{z^{(N)}}||<\epsilon$. 
If $\lambda^v_i=0$ for all $v\in V(\gamma_i)$ then $T_i$ is a zero 
volume eigenstate for all $V(O)$ and we may pick $T_i+\epsilon T_z$
where $T_z$ is in the cone and explicitly given by a state of the form 
(\ref{a.16}) with only positive $z_i$. 
If $\lambda^v_i>0$ for at least one $v_0\in V(\gamma_i)$
we set $T_0=T_{i_0}:=T_i$ and $T_n:=T_{i_n}$ 
to be the parallel translate of $T_0$ by 
the translation translating $v_0$ to $v_n$ where 
$\mathbb{N}_0\to \mathbb{Q}^3,\; n\mapsto v_n$ 
is a bijection. Let $z^{(N)}_n$ for
$0\le n,N\le M$ be a non-degenerate $(M+1)\times (M+1)$ matrix 
$A^N\;_n$ with only positive entries, for instance a Van der Monde 
matrix $A_n\;^N=[x_n]^N,\;0<x_0<x_1<..<x_M$ with 
$\det(A)=\prod_{0\le m<n\le M} \;[x_n-x_m]$ \cite{31}. Let 
$c_N=[A^{-1}]_N\;^0$ then 
\be \label{a.21}
\sum_{N=0}^\infty c_N \; T_{z^{(N)}} -T_i
=[\sum_{N=0}^M c_N \;\sum_{n=0}^M\; z^{(N)}_n\; T_n -T_0]
+\sum_{N=0}^M c_N \;\sum_{n=M+1}^\infty\; z^{(N)}_n\; T_n 
=\sum_{N=0}^M c_N \;\sum_{n=M+1}^\infty\; z^{(N)}_n\; T_n 
\ee
Accordingly 
\ba \label{a.22}
&&||\sum_{N=0}^M c_N \; T_{z^{(N)}} -T_i||^2
=\sum_{n=M+1}^\infty\;|\sum_{N=0}^M c_N\;z^{(N)}_n|^2
\nonumber\\
&\le & \sum_{n=M+1}^\infty\;[\sum_{N=0}^M |c_N|^2]\;
[\sum_{N=0}^M\; |z^{(N)}_n|^2]
\ea
by the CS inequality. Pick for $n>M,\;0\le N\le M$
\be \label{a.23}
z^{(N)}_n=\frac{\kappa\;\epsilon}{\sqrt{M}\;n},\;\;
\kappa^{-2}=[\sum_{N=0}^M\; |c_N|^2]\;[\sum_{n=M+1}^\infty\; n^{-2}]
\ee
Then $||T_i-\sum_{N=0}^M c_N\; T_{z^{(N)}}||<\epsilon$ and each 
$T_{z^{(N)}},\; N=1,..,M$ is a linear combination of the $T_n$ with 
only positive coefficients. Moreover 
\be \label{a.24}
<T_{z^{(N)}}, V(O)\; T_{z^{(N)}}>=\sum_{n=0}^\infty
[z^{(N)}_n]^2\; \Lambda(O,i_n)
\ee
which is bounded from above by $||T_{z^{(N)}}||^2\;
\sum_{v\in V(\gamma_i)} \lambda^v_i$ and is strictly larger than zero 
for any $O$. A similar upper bound holds for $V(O)$ replaced by $V(O)^2$.\\
iv.\\
Let $T_z\in {\cal C}_{{\rm ND}}$. Pick a phase $e_i, |e_i|=1$ for each 
$i\in I$ and set $z^e_i:=z_i \; e_i$. Then 
$T_{z^e}\in {\cal M}_{{\rm ND}}$ with 
$<T_{z^e},\;V(O)^k\;T_{z^e}>=<T_z,(V(O))^k T_z>,\;k=0,1,2$ but 
$T_{z^e}\not \in {\cal C}_{{\rm ND}}$ as soon as one of $e_i\not=1$. 
\end{proof}
Since ${\cal C}_{{\rm ND}}$ is a cone one cannot obviously generate 
it from elements $E_I\in {\cal C}_{{\rm ND}}$ by taking finite 
linear combinations such that $E_I$ is also an ONB of $\cal H$. We have 
\be \label{a.25}
<T_z,T_{z'}>=\sum_{i\in I\cap I'} \; z_i\; z'_i
\ee 
i.e. in order that $T_z\;\perp\; T_{z'}$ we need that $I\cap I'=\emptyset$
while $I(O)\cap I,I(O)\cap I'\not=\emptyset$ for any open $O$. This is 
certainly possible using the above translate construction for a 
countable number of index sets $I$, for one can construct a countable 
number of countable sets of points which are mutually disjoint and 
dense (e.g. use $v^{(p)}_{N,n,k}=N+\frac{k}{p^n}$ where $p$ is a prime and 
$1\le k\le p^n-1$ is relative prime to $p$ i.e. $k\not=lp$. Then 
$v^{(p)}_{N,n,k}=v^{(p')}_{N',n',k'}$ requires $N=N'$ and 
$k' p^n=k (p')^{n'}$ but for $p\not=p'$ both primes this is impossible). 
 
\subsection{Non-degenerate distributions}
\label{sa.4}

The previous subsection has revealed the importance of the 
cone ${\cal C}_{{\rm ND}}$ in $\cal H$. Its members are elements of 
$\cal H$ and thus by the Riesz lemma define continuous linear functionals 
\be \label{a.26}
l=<T_z,.>=\sum_{i\in {\cal I}} z_i\; <T_i,.>
\ee
with $z_i\ge 0$ and $z_i>0$ for $i$ in countable $I\subset {\cal I}$
such that $\sum_i z_i^2<\infty$ and $I\cap I(O)\not\emptyset$ for 
all open $O$. Following the strategy to solve the 
constraints in LQG and the habitat construction (reviewed 
e.g. in \cite{13})
it is thus motivated to drop the condition $\sum_i z_i^2<\infty$ and
just keep $z_i\ge 0$ and
$I\cap I(O)\not=\emptyset$ for 
all open $O$. Then (\ref{a.26}) becomes a distribution, i.e.
an element of the 
algebraic dual $L_{{\rm ND}}$ if we further constrain the $z_i$ for 
$T_z\in {\cal C}_{{\rm ND}}$ to be of ``rapid dcrease'' in $i\in {\cal I}$.
Intuitively this means that if we label the $i\in I$ by $n\in \mathbb{N}$ 
then the 
decay in $n$ is stronger than any power in $n$ which allows the $z_i$ for 
$T_z\in L_{{\rm ND}}$ to grow polynomially in $i$.
\begin{Definition} \label{defa.3} ~\\
i.\\
The net $(z_i)_{i\in {\cal I}}$ is called of rapid decrease iff 1. it has 
countable support $I\subset {\cal I}$ and 2. there eixists a bijection
$b:\; \mathbb{N}\to I$ such that for every $N\in \mathbb{N}_0$
\be \label{a.27}
||z||_N:=\sup_{n\in \mathbb{N}}\; |n^N z_{b(n)}|\;<\infty
\ee
ii.\\
The Schwartz subspace ${\cal S}$ of $\cal H$ consists of those 
$T_z=\sum_{i\in {\cal I}}\; z_i\; T_i$ such that $z$ is of rapid 
decrease. Its intersection with ${\cal M}_{{\rm ND}}$ is denoted 
as ${\cal S}_{{\rm ND}}$. $\cal S$ carries the subspace topology induced 
from $\cal H$.\\
iii.\\
The (continuous) linear functionals on $\cal S$, also called the 
(topological) algebraic dual, is denoted by  (${\cal S}'$) ${\cal S}^\ast$.
iv.\\
For an element $l$ in the (topological) algebraic dual we define 
its support $I$ to consist of those $i\in {\cal I}$ such that 
$l(T_i)\not=0$ and for each 
open $O$ the set $I(O)\subset {\cal I}$ as the set of $i$ such that 
$l(V(O)\;T_i)\not=0$. 
Such an element is called non-degenerate 
iff $I(O)\not=\emptyset$ for all $O\in {\cal O}$.
We denote the set of these distributions as 
(${\cal S}_{{\rm ND}}'$) ${\cal S}_{{\rm ND}}^\ast$.
\end{Definition}
Note that algebraic distributions can have uncountable support and 
have no growth conditions on the $z_i:=l(T_i)$. In fact, there can 
be none because for uncountable support the rapid decrease definition 
cannot even be formulated. 
We expect continous distributions to have countable support  
but much milder growth conditions on the $z_i=l(T_i)$ than the Schwartz 
space vectors have. Note that the rapid decrease condition implies 
that with respect to the labelling of single non zero volume 
vertex states with vertices at  
$v_{N,n,k}=N+\frac{k}{2^n}$ of subsection \ref{sa.2} there 
is a bijection $b:\;m\mapsto N(m),n(m),k(m)$ with $k(m)\in M_{n(m)}$ 
and $z_{b(m)}$ decays rapidly with respect to $m$. Hence the superposition 
of VNWF $T_z$ is controlled both in zooming out indefinitely $||N||\to
\infty$ and in zooming in $n\to \infty$ indefinitely. Thus we are 
controlling both the macroscopic and microscopic realm i.e. the long 
and short distance behaviour and in that sense the analogy with the 
Schwartz function topology on $\mathbb{R}$ is very close because 
the control of the derivative suprema of Schwartz functions is a 
short distance condition. For more than one vertex,  
consider e.g. for each $M\in \mathbb{N}$ a function 
$z_M:\; \sigma^M\to \mathbb{C}$ symmetric 
under permutation of arguments (otherwise some ordering prescription
is required) and $z_i:=z_{|V(\gamma_i)|}(\{v\}_{v\in V(\gamma_i)})$. 
For elements in ${\cal S}$ one has to restrict $i$ to a countable subset,
for algebraic distributions this is not the case.\\
\\
As an example for a physically interesting 
non degenerate distribution and in view of the prominent role of the 
volume operator in quantum gravity, 
complexifier coherent states \cite{22} based  
on the total volume $V:=V(\sigma)$ appear natural. It induces the natural
complexification the classical configuration space of connections given 
by $Z_a^j=A_a^j-i\; e_a^j,\;e_a^j=\{V,A_a^j\}$ being the co-triad. The
$Z_a^j$ separate the points $A_a^j,E^a_j$ iff 
$q_{ab}=e^j_a e^k_b\delta_{jk}$ is non-degenerate since 
$E^a_j=\sqrt{\det(q)} q^{ab} e_b^j$ which again underlines the 
importance of the non-degeneracy condition.
\begin{Definition} \label{defa.3} ~\\
Let $V$ be the total volume. The volume complexifier coherent distribution is 
defined as \cite{22}
\be \label{a.28}
l_Z=[e^{-V/L^3} \delta]_{A\to Z}
\ee
where $L$ is a parameter of length dimension and $A\to Z$ denotes 
analytic extension.
\end{Definition}
We can work out the ``Fourier coefficients'' of $l_Z$ with respect 
to the ONB $T_i$ explicitly
\be \label{a.29}
l_Z(T_i)=T_i(Z)\;e^{-\Lambda(\sigma,i)/L^3}
\ee
where $T_i(Z)$ is the analytic extension of $T_i(A)$ (which is a 
polynomial in the holonomies of the classical connection $A$ along 
the edges of $\gamma_i$). The closed formula for (\ref{a.28}) is thus  
\be \label{a.30}
l_Z=\sum_i \; l_Z(T_i)\; <T_i,.>
\ee
Thus while 
(\ref{a.29}) is well defined for each $i$,
$l_Z$ has uncountably infinite support.
Moreover, there are uncountably many $i$ with $\Lambda(\sigma,i)=0$ so that 
the growth of $l_Z(T_i)$ in $i$ 
is not suppressed by the exponential in (\ref{a.29}). Even if 
$V(\sigma_i)>0$, it maybe that $V(\sigma,i)$ has ``flat directions'' 
with respect to the spin labels $j_e$ while $T_i(Z)$ is known to have 
exponential growth with respect to the spin label if $\Im(Z)\not=0$. 
Accordinly, $l_Z$ is in the algebraic dual and not in the topological
dual. On the other hand, $\Lambda(O,i)\not=\emptyset$ for all $O$ thus 
$l_Z$ is non degenerate for all $Z$.

Thus, for the $l_Z$ with $Z$ varying over the classical phase space 
to be useful as a habitat one presumably has to restrict the sum 
in (\ref{a.30}) to a countable non-degenerate subset $I$ without 
violating the non-degeneracy condition. Another important 
issue of practical concern is that knowlede of $V(O,i)$ for such $i$
to sufficient accuracy is required. Given that these conditions can 
be met, one might be able to restrict $l_Z$ to an element $l'_Z$ in the 
topological dual such that the quantisations of the classical 
regulated constraints with density weight 
unity act dually as  
\be \label{a.31}
C'_\epsilon(n)\;l'_Z=i\;l'_{X_{C_\epsilon(n)}\cdot Z},\; 
D'_\epsilon(u)\;l'_Z=i\;l'_{X_{D_\epsilon(u)}\cdot Z},\; 
\ee
plus subleading corrections in $\epsilon$
where $X_F$ denotes the Hamiltonian vector field of a phase space function 
$F$. Then one could take the limit $\epsilon\to 0$ (pointwise wrt 
the $T_i$) and obtain in the optimal case 
\be \label{a.32}
C'(n)\;l'_Z=i\; l'_{X_{C(n)}\cdot Z},\; 
D'(u)\;l'_Z=i\; l'_{X_{D(u)}\cdot Z},\; 
\ee
with the classical continuum constraints $C(n),D(u)$ 
which would grant anomaly closure on this habitat. \\
\\
Such a calculation is presently out of reach because the spectrum 
of the volume operator is presently not known in sufficient detail.
This is one of the reasons for why we are interested in solving the
Gauss constraint already classically thereby opening access to representations 
in which the volume operator spectrum is available in closed form as we have 
shown in section \ref{s5}, see also \cite{32}.
In the next subsection we will therefore limit ourselves to
carry out a preliminary investigation  
in the spirit of (\ref{a.31})
which actually
takes place entirely within $\cal H$, i.e. it does not refer to any habitats.
However, still the concept of non-deneracy will be crucial.

\subsection{Preliminary investigation of the action
of $\mathfrak{h}$ on non-degenerate LQG states}
\label{sa.5}

Here we just sketch general properties and contrast 
them with those earlier analyses \cite{22}:\\
\\
I. Technical differences:\\
A.\\ 
Coherent states in the past were constructed on a {\it single}  
graph $\gamma$ 
which admits a dual cell complex $\gamma^\ast$ and which has the property 
that for each open region $O$ with compact closure the set 
$V(\gamma)\cap O$ is finite. Thus while one can consider an infinite 
graph in what is called the infinite tensor product extension of LQG
\cite{22}, the set $V(\gamma)$ 
is locally finite. Such a state is therefore {\it degenerate}.

By contrast, here one would consider 
a {\it countably infinite set of finite graphs $\gamma_i$} which admit 
dual cell complexes $\gamma_i^\ast$. 
While each $\gamma_i$ is a finite graph, its union 
$\cup_i V(\gamma_i)$
at which the volume of the 
superposition is excited is not at all locally finite but 
rather fills all of $\sigma$ densely. The state is therefore 
{\it non-degenerate}.\\
B.\\
An infinite tensor product is no longer in the LQG Hilbert space
but is a normalisable state in a genuine extension of the 
LQG Hilbert space which 
contains the LQG Hilbert space as one of an uncountable 
number of mutually orthogonal subspaces. The 
infinite superpositions discussed above are still in the LQG Hilbert 
space even if its vertices fill all of $\sigma$ densely.\\
C.\\
In these earlier works the coherent states were {\it (in)finite 
tensor products} 
of states where the {\it factors} are labelled by the edges of $\gamma$. 
By contrast, here one would consider {\it infinite superpositions} of states 
where the {\it summands} are labelled by a vertex of the graphs $\gamma_i$.\\
D.\\
In \cite{22} the classical degrees freedom whose fluctuations are suppressed 
within the Heisenberg uncertainty bound 
were holonomies along the edges of $\gamma$ and 
fluxes though the faces of $\gamma^\ast$. By contrast, in the construction 
here the semiclassical degrees of freedom are holonomies along edges of 
the $\gamma_i$ and faces of the $\gamma_i^\ast$.\\
\\
II. Conceptual difference\\
In previous semiclassical analyses of the hypersurface deformation 
algebra $\mathfrak{h}$ 
\cite{26} on a single graph $\gamma$ one did {\it choose $\gamma$ to fill
$\sigma$ sufficiently densely} so that exectation values approximate 
Riemann sums and thus classical integrals. However, no reason 
was given why this computation should be an indication of anomaly freeness.
After all, while coherent states are overcomplete on a given graph
those states are far from complete in the full Hilbert space which 
concerns all graphs 
and most of these graphs do not fill $\sigma$ sufficiently densely
(in particular not if $\sigma$ is not compact). 
Therefore, if one asked for anomaly freeness on the full Hilbert space 
then this computation on a single graph is insufficient.

The new input is that we ask anomaly freeness only on non-degenerate states
because we consider non-degenerate states as the only sector on which 
$\mathfrak{h}$ should be checked. 
Therefore anomaly freeness has to be checked
on a much smaller set of states and and for each of them 
the union of the underlying 
$\gamma_i$ {\it automatically fills $\sigma$ densely}.\\
\\
III. Dynamical stability of non-degeneracy\\
As is well known, the action of the classical constraints on $V(O)$ is 
\be \label{a.33}
\{D(u),V(O)\}=[\frac{d}{ds} V(\varphi^u_s(O))]_{s=0},\;\;
\{C(n),V(O)\}=\frac{1}{2}\;\int_O\;dx\;n\; q^{ab}\; K_{ab}
\ee
where $K$ is the extrinsic curvature. Thus spatial diffeomorphisms preserve 
non-degeneracy, temporal diffemorphisms not necessarily (c.f. collapsing 
cosmological solutions). In the quantum theory we are interested in the 
expansion 
\be \label{a.34}
C(n)\; T_z=T_{z'},\; z'_i=\sum_j\; z_j \; <T_i, C(n) T_j>
\ee
and the corresponding $I', I'(O)\subset I'$, given that $I(O)\subset I$
is not empty for any $O$. Due to the local action of $C(n)$ we we know  
that $C(n) T_j$ is non vanishing only if $V(\gamma_j)\subset {\rm supp}(n)$
and thus $I'(O)=\emptyset$ for $O\cap$supp$(n)=\emptyset$. Thus we can 
have preservation of quantum non-degeneracy only if supp$(n)=\sigma$ which 
is expected taking the assumed don-degeneracy of the classical spacetime 
metric into account. If that condition is met, then, given 
(\ref{a.34}), $I'(O)\not=\emptyset$
appears to be satisfield for generic $I(O)\not=\emptyset$ because if 
$I(O)\not=\emptyset$ then $I(O)$ is a countably infinite set as $O$ is 
open and $\cup_{i\in I} V(\gamma_i)$ is dense. Thus, in order to have
e.g. 
$z'_i=0$ for all $i\in I(O)$ all the countably infinite 
number of matrix elements $<T_i, C(n) T_j>,\;
i,j\in I(O)$ would have to vanish, which, given the explicit form 
of $C(n)$ is not very likely for generic $I(O)$. Thus preservation 
of quantum degeneracy at least under finitely many actions of $C(n)$ 
with $n$ smooth and nowhere vanishing (but maybe of rapid decrease) 
appears to be rather likely.

\end{appendix}


\begin{thebibliography}{99}

\parskip -5pt

\bibitem{1} E. Cartan. The theory of spinors. Dover Publications Inc., 
Dover, 2003

%\bibitem{2} Palatini

\bibitem{3} S. Holst.
Barbero's Hamiltonian derived from a generalized Hilbert-Palatini action.
Phys. Rev. {\bf D 53} (1996) 5966-5969, e-Print: gr-qc/9511026 [gr-qc]

\bibitem{4} C. Rovelli. Quantum Gravity. Cambridge University
Press, Cambridge, 2004.\\
T. Thiemann. Modern Canonical Quantum General Relativity. Cambridge
University Press, Cambridge, 2007\\
J. Pullin, R. Gambini. A first course in Loop Quantum Gravity.
Oxford University Press, New York, 2011\\
C. Rovelli, F. Vidotto. Covariant Loop Quantum Gravity. Cambridge
University Press, Cambridge, 2015

\bibitem{5} A. Ashtekar, C.J. Isham. Representations of the Holonomy
Algebras of Gravity and Non-Abelean Gauge Theories.
Class. Quantum Grav. {\bf 9} (1992) 1433, [hep-th/9202053]\\
A. Ashtekar, J. Lewandowski. Representation
theory of analytic Holonomy $C^\star$ algebras. In: Knots and
Quantum Gravity, J. Baez (ed.), Oxford University Press, Oxford 1994\\
A. Ashtekar, J. Lewandowski. Projective Techniques and
Functional Integration for Gauge Theories. J.
Math. Phys. {\bf 36}, 2170 (1995), [gr-qc/9411046]\\
C. Fleischhack. Representations of the Weyl algebra in quantum geometry.
Commun. Math. Phys. {\bf 285} (2009) 67-140, [math-ph/0407006]\\
J. Lewandowski, A. Okolow, H. Sahlmann, T. Thiemann.
Uniqueness of diffeomorphism invariant states on holonomy-flux algebras.
Commun. Math. Phys. {\bf 267} (2006) 703-733, [gr-qc/0504147]

\bibitem{6} A. Ashtekar. New Variables for Classical and Quantum Gravity.
Phys. Rev. Lett. {\bf 57} (1986) 2244-2247\\
J. F. G. Barbero, A real polynomial formulation of general relativity in
terms of connections, Phys. Rev. {\bf D49} (1994) 6935-6938

\bibitem{7}  M. Creutz. Quarks, Gluons and Lattices.
Cambridge University Press, Cambridge, 1985.

\bibitem{8} C. Rovelli and L. Smolin.
Spin networks and quantum gravity. Phys. Rev. {\bf D52} (1995) 5743-5759;
e-Print: gr-qc/9505006 [gr-qc]

\bibitem{9} A. Ashtekar, A. Corichi, J.-A. Zapata 

Quantum theory of geometry III: Noncommutativity of Riemannian structures.
Class. Quant. Grav. {\bf 15} (1998) 2955-2972, e-Print: gr-qc/9806041 [gr-qc]

\bibitem{10} C. Rovelli and L. Smolin.
Discreteness of volume and area in quantum gravity.
Nucl. Phys. {\bf B442} (1995), 593-622; Erratum: Nucl. Phys.
{\bf B456} (1995) 753, [gr-qc/9411005]\\
A. Ashtekar and J. Lewandowski.
Quantum theory of geometry I: Area Operators.
Class. Quant. Grav. {\bf 14} (1997) A55-A82, [gr-qc/9602046];
Quantum theory of geometry II:
Volume operators. Adv. Theo. Math. Phys. {\bf 1} (1997) 388-429,
[gr-qc/9711031]

\bibitem{11} T. Thiemann. Anomaly-free Formulation of non-perturbative,
four-dimensional Lorentzian Quantum Gravity. Physics Letters {\bf B380}
(1996) 257-264, [gr-qc/9606088]\\
T. Thiemann. Quantum Spin Dynamics (QSD).
Class. Quantum Grav. {\bf 15} (1998) 839-73, [gr-qc/9606089];
Quantum Spin Dynamics (QSD) : V.
Quantum Gravity as the Natural Regulator of the Hamiltonian Constraint
of Matter Quantum Field Theories.
Class. Quantum Grav. {\bf 15} (1998) 1281-1314, [gr-qc/9705019]

\bibitem{12} J. D. Brown, K. V. Kuchar.
Dust as a standard of space and time in canonical quantum gravity.
Phys. Rev. {\bf D51} (1995) 5600-5629.[gr-qc/9409001]\\
K. V. Kuchar, C. G. Torre,
Gaussian reference fluid and interpretation of quantum geometrodynamics.
Phys. Rev. {\bf D43} (1991) 419-441.\\
V. Husain, T. Pawlowski. Time and a physical Hamiltonian for quantum gravity.
Phys.Rev.Lett. 108 (2012) 141301. e-Print: 1108.1145 [gr-qc]\\
M. Domagala, K. Giesel, W. Kaminski, J. Lewandowski.
Gravity quantized: Loop Quantum Gravity with a Scalar Field.
Phys. Rev. {\bf D82} (2010) 104038, [arXiv:1009.2445]\\
K. Giesel, T. Thiemann. Scalar Material Reference Systems and Loop Quantum
Gravity. Class. Quant. Grav. {\bf 32} (2015) 135015,
[arXiv:1206.3807]

\bibitem{13} T. Thiemann.
Non-degenerate metrics, hypersurface deformation algebra,
non-anomalous representations and density weights in quantum gravity.
e-Print: 2207.08299 [gr-qc]

\bibitem{13a} E. Alesci, F. Cianfrani.
Quantum Reduced Loop Gravity and the foundation of Loop Quantum Cosmology.
Int. J. Mod. Phys. {\bf D 25} (2016) 08, 1642005, e-Print: 1602.05475 [gr-qc]

\bibitem{14} T. Jacobson. Fermions in Canonical Gravity.
Class. Quant. Grav. {\bf 5} (1988) L143


\bibitem{15} R. M. Wald. General Relativity. The University of
Chicago Press, Chicago, 1989

\bibitem{16} N. Bodendorfer, T. Thiemann, A. Thurn.
New variables for classical and quantum (super)-gravity in all dimensions.
PoS QGQGS2011 (2011) 022\\
N. Bodendorfer, T. Thiemann, A. Thurn.
New Variables for Classical and Quantum Gravity in all Dimensions
I. Hamiltonian Analysis. Class. Quant. Grav. {\bf 30} (2013) 045001, e-Print:
1105.3703 [gr-qc];
II. Lagrangian Analysis. Class. Quant. Grav. {\bf 30} (2013) 045002, e-Print:
1105.3704 [gr-qc]; III. Quantum Theory.
Class. Quant. Grav. {\bf 30} (2013) 045003, e-Print: 1105.3705 [gr-qc]

%\bibitem{17} spin structure

\bibitem{18} M. Henneaux, C. Teitelboim. Quantisation of Gauge
Systems. Princeton University Press, Princeton, 1992

\bibitem{19} P.A.M Dirac, Phys. Rev. {\bf 73} (1948) 1092;
Rev. Mod. Phys. {\bf 21} (1949) 392\\
J. A. Wheeler. Geometrodynamics. Academic Press, New York, 1962\\
B. S. DeWitt, Phys. Rev. {\bf 160} (1967) 1113; Phys. Rev.
{\bf 162} (1967) 1195; Phys. Rev. {\bf 162} (1967) 1239.



\bibitem{20} T. Thiemann
Reality conditions inducing transforms for quantum gauge field theory and 
quantum gravity. Class. Quant. Grav. {\bf 13} (1996) 1383-1404, e-Print:
gr-qc/9511057 [gr-qc]\\
M. Varadarajan.
From Euclidean to Lorentzian Loop Quantum Gravity via a Positive Complexifier
Class. Quant. Grav. {\bf 36} (2019) 1, 015016, e-Print: 1808.00673 [gr-qc]

\bibitem{21} C. Itzykson, J.-B. Zuber. Quantum field Theory.
Dover books on physics, Dover, 2006

\bibitem{22} H. Narnhofer, W.E. Thirring.
Covariant QED without indefinite metric.
Rev. Math. Phys. {\bf 4} (1992) spec01, 197-211

\bibitem{23} S. Eilenberg, N. Steenrod (1952-12-31). Foundations 
of Algebraic Topology. Princeton University Press, Princeton, 2015

%\bibitem{24} AL area and volume ops

\bibitem{25} E. Alesci, J. Lewandowkski, M. Assanioussi
Curvature operator for loop quantum gravity. 
Phys. Rev. {\bf D 89} (2014) 12, 124017, e-Print: 1403.3190 [gr-qc]

\bibitem{26} T. Thiemann, O. Winkler.
Gauge field theory coherent states (GCS) 4: 
Infinite tensor product and thermodynamical limit
Class. Quant. Grav. {\bf 18} (2001) 4997-5054, e-Print:
hep-th/0005235 [hep-th]

%\bibitem{27} TT non deg states

\bibitem{31} R. A. Horn, C. R. Johnson. Topics in matrix analysis. 
Cambridge University Press, Cambridge, 1991.

\bibitem{32} T. Thiemann. Exact quantisation of $U(1)^3$ 
quantum gravity via exponentiation of the hypersurface deformation algebroid.
e-Print: 2207.08302 [gr-qc]

\bibitem{33} T. Thiemann. Properties of a smooth, dense, invariant domain
for singular potential Schr\"odinger operators. 

\bibitem{34} K. Giesel, J. Tambornino, T. Thiemann. Born-Oppenheimer 
decomposition for quantum fields on quantum spacetimes. 
e-Print: 0911.5331 [gr-qc] 

\bibitem{35} S. Schander, T. Thiemann. Backreaction in Cosmology.
Front. Astron. Space Sci. {\bf 0} (2021) 113, 
Front. Astron. Space Sci. {\bf 8} (2021) 692198. e-Print: 2106.06043 [gr-qc]

\bibitem{36} I. Agullo, P. Singh. Loop Quantum Cosmology.
e-Print: 1612.01236 [gr-qc]

\bibitem{37} T. Thiemann. On the fate of the classical big bang singularity 
in the Schr\"odinger representation. In preparation.

\end{thebibliography}
\end{document}